\documentstyle[aps,pre,eqsecnum,epsf]{revtex}
\begin{document}
 \renewcommand{\thesection}{\arabic{section}}
 \twocolumn[{
 \draft
 \title{Exact  Resummations in the Theory of Hydrodynamic Turbulence:\\
 II.  A Ladder to Anomalous Scaling}
 \author {Victor L'vov\cite{lvov}  and Itamar  Procaccia\cite{procaccia}  }
 \address{Departments of~~\cite{lvov}Physics of Complex Systems
 {\rm and}~~\cite{procaccia}Chemical Physics,\\
  The Weizmann Institute of Science,
 Rehovot 76100, Israel,\\
 \cite{lvov}Institute of Automation and Electrometry,
  Ac. Sci. of Russia, 630090, Novosibirsk, Russia
   }
 \maketitle
 \widetext
 \begin{abstract}
 \leftskip 54.8pt
 \rightskip 54.8pt
 In paper I of this series on fluid turbulence we showed that exact
 resummations of the perturbative theory of the structure functions of
 velocity differences result in a finite (order by order) theory. These
 findings exclude any known perturbative mechanism for anomalous scaling of
 the velocity structure functions. In this paper we continue to build the
 theory of turbulence and commence the analysis of nonperturbative effects
 that form the analytic basis of anomalous scaling. Starting from the
 Navier-Stokes equations (at high Reynolds number Re) we discuss the
 simplest examples of the appearance of anomalous exponents in fluid
 mechanics. These examples are the nonlinear (four-point) Green's function
 and related quantities. We show that the renormalized perturbation theory
 for these functions contains ``ladder`` diagrams with (convergent!)
 logarithmic terms that sum up to anomalous exponents. Using a new  sum
 rule which is derived here we calculate the leading anomalous exponent and
 show that it is critical in a sense made precise below. This result opens
 up the possibility of multiscaling of the structure functions with the
 outer scale of turbulence as the renormalization length. This possibility
 will be discussed in detail in the concluding paper III of this series.
 \end{abstract}
 \leftskip 54.8pt

 }]
 \narrowtext
  \section{Introduction}
 In this paper we clarify, on the basis of an analytic theory, how
 anomalous scaling appears in fluid mechanics. The aim of the analytic
 theory is to reach understanding on the basis of the Navier-Stokes
 equations. We thus make a clear cut distinction between the analytical
 approach and the host of ad hoc models that were employed to attempt a
 description of anomalous scaling in turbulence.

 The common wisdom about anomalous scaling in turbulence can be found in a
 number of recent reviews  and books (see, e.g. \cite{95Nel,Fri}). In
 terms of the scaling exponents, most attention was given to those
 associated with the structure functions of velocity differences and to the
 correlation function of the fluctuations in the rate of energy
 dissipation. In terms of the Eulerian velocity field ${\bf u}({\bf r},t)$
 the structure functions $S_{n}(R)$ can be defined\cite{MY-2} in terms of
 the velocity differences $\delta {\bf u}( {\bf r}+ {\bf R}, {\bf r},t)$:

 \begin{eqnarray}
 \delta{\bf u}( {\bf r}+ {\bf R}, {\bf r},t) &\equiv&
 [ {\bf u}( {\bf r}+ {\bf R},t)- {\bf u}( {\bf r},t)]\, ,
 \label{a1} \\
  S_{n}(R)&\equiv & \langle | \delta {\bf u}( {\bf r}+ {\bf R}, {\bf
 r},t)| ^{n}\rangle
 \label{a2}
 \end{eqnarray}
 where the symbol $\langle\dots\rangle$ denotes an average over time and
 over {\bf r}. The structure functions $S_{n}(R)$ are expected to exhibit
 scaling behavior for values of $R$ in the ``inertial range" $L\gg R\gg
 \eta$ where $L$ and $\eta$ are the outer scale and the Kolmogorov
 dissipation scale respectively:
 \begin{equation}
  S _{n}(R) \sim C R ^{\zeta_n}\   .
 \label{a3}
 \end{equation}
 In the Kolmogorov theory (K41) the exponents $\zeta_{n}$ equal $n/3$,
 whereas experiments and popular belief ascribed anomalous values to these
 exponents. The energy dissipation rate ${\varepsilon }( {\bf r},t)$ is the
 field
 \begin{equation}
 {\varepsilon }( {\bf r},t) \equiv {\nu\over 2}
 \Big[\partial _{\alpha}u_{\beta}( {\bf r},t)+\partial_{\beta} u _{\alpha}
 ( {\bf r},t)\Big]^{2}
 \label{a4}
 \end{equation}
 where $\nu$ is the kinematic viscosity. The ``centered" correlation
 function of the fluctuations in this field, $K _{\varepsilon\varepsilon}
 (R)$, is defined as
 \begin{equation}
 K_{\varepsilon\varepsilon}(R) =  \langle {\hat\varepsilon}( {\bf r}
 + {\bf R},t) {\hat\varepsilon}( {\bf r},t)\rangle\,,
 \label{a5}
 \end{equation}
 where  $\hat {\varepsilon}( {\bf r},t) = {\varepsilon }( {\bf r},t)  -
 \bar{\varepsilon}$, and $\bar{\varepsilon}$ is the mean of the dissipation
 field.  It was found in experiments that $K_{\varepsilon \varepsilon}(R)$
 decays very slowly in the inertial range,
 \begin{equation}
 K_{\varepsilon\varepsilon} (R)  \sim  R ^{-\mu}
 \label{a6}
 \end{equation}
 with $\mu$ having a numerical value\cite{93SK} in the range $0.25\pm 0.05$.
 It was claimed\cite{MY-2} that the K41 theory required $\mu$ to  vanish.
 Accordingly, there have been many attempts to construct models of
 turbulence (see, i.e. \cite{95Nel,Fri,MY-2,62Kol,64NS,74Man,78FSN,85PF}
 and references therein) to take (\ref{a6}) into account and to explain how
 measured deviations $(\zeta_{n} -n/3)$ in the exponents of the structure
 functions were related to $\mu$.

 In this paper we show that the renormalized perturbation theory for
 correlation functions that include velocity  derivatives (to second or
 higher power) exhibit in their perturbation expansion a logarithmic
 dependence on the  viscous scale $\eta$ \cite{95LL,94LL}. In this way the
 inner scale of turbulence appears explicitly in the analytic theory. The
 perturbative series can be resummed to obtain integrodifferential
 equations for some many-point objects of the theory. These equations have
 also non-perturbative scale-invariant solutions that can be represented as
 power laws of $\eta$ to some exponents $\Delta$. For example it will be
 shown in paper III that the correlation function of the energy dissipation
 field has such dependence:
 \begin{equation}
 K_{\epsilon\epsilon}(R) \sim \bar\epsilon^2 (L/R)
^{2\zeta_2-\zeta_4} (R/\eta)^
 {2(\Delta-\Delta_c)}
 \label{Kee1}
 \end{equation}
 where $\Delta_c=2 - \zeta_2$. It has been argued \cite{95LP} that if
 $\Delta<\Delta_c$ (a situation referred to as the ``subcritical
 scenario"), then K41 is asymptotically exact for infinite Re. Then
 $2\zeta_2=\zeta_4 = 4/3$ and the outer scale $L$ disappears from
 (\ref{Kee1}). In that case the exponent $\mu$ is identified with
 $2(4/3-\Delta)$, and the renormalization length is the inner length
 $\eta$. In fact, it will be shown here that the exponent $\Delta$ can be
 computed explicitly, and that it takes on exactly the value $\Delta =
 \Delta_c$. As a result of this the correlation $K_{\epsilon\epsilon}(R)$
 can be shown to depend on $R$ like
 \begin{equation}
 K_{\epsilon\epsilon}(R) =\sim \bar\epsilon^2 (R/L )^{\zeta_4-2\zeta_2}
 \ .\label{Kee2}
 \end{equation}
 In other words, the critical situation $\Delta = \Delta_c$ results in the
 disappearance of the inner renormalization scale and the appearance of the
 outer renormalization scale in (\ref{Kee2}). In addition one notes that
 $K_{\epsilon\epsilon}(R)$ decays as a function of $R$ (i.e. the
 correlation is mixing) only if $\zeta_4 < 2\zeta_2$  which implies
 deviations from K41. Thus we will argue in paper III that the critical
 scenario $\Delta = \Delta_c$ goes hand in hand with multiscaling if
 $K_{\epsilon\epsilon}(R)$ is mixing and then $\mu$ is identified with
 $\zeta_4 - 2\zeta_2$.

 The quantity that displays the existence of the anomalous exponent
 $\Delta$ in the simplest possible way is the nonlinear (4-point) Green's
 function. Thus, after reviewing in Section 2 some essential results from
 previous work, we turn in Section 3 to the nonlinear Green's function. We
 develop its diagrammatic representation and show how all the diagrams in
 its expansion can be resummed exactly into ladder diagrams. We discuss the
 properties of the ladder diagrams, and demonstrate in Section 4 that the
 dressing of the ladder diagrams does not change the fact that they resum
 to give power laws with anomalous exponents. In section 5 we firstly
 demonstrate that the resummed equation for the Nonlinear Green's function
 indeed has a nonperturbative solution with an anomalous exponent $\Delta$,
 and then derive a sum rule that allows us to conclude that $\Delta =
 \Delta_c =2-\zeta_2$. Last Section 6 concludes this paper.
 \section{ Summary of previous results }
 \label{19-sect2}
\begin{figure}
 \epsfxsize=6truecm
 \epsfbox{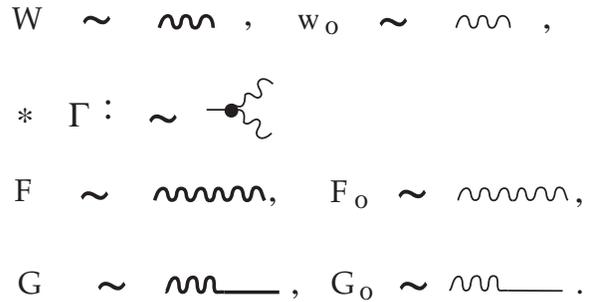}
 \vspace{.5cm}
 \caption{ Basic graphical symbols for ${\bf w}$, BL velocity differences
 (2.4);  $\Gamma$, vertex -- amplitude of interaction; $F$ double
 correlation function of the  BL velocity differences  (2.7); $G$, Green's
 function (2.6) ${\bf w}_0$, $F_0$ and $G_0$, corresponding values  in
 zero order approximation with respect to interaction (``bare values"). }
 \label{fig:fig1}
 \end{figure}

\begin{figure}
 \epsfxsize=8.6truecm
 \epsfbox{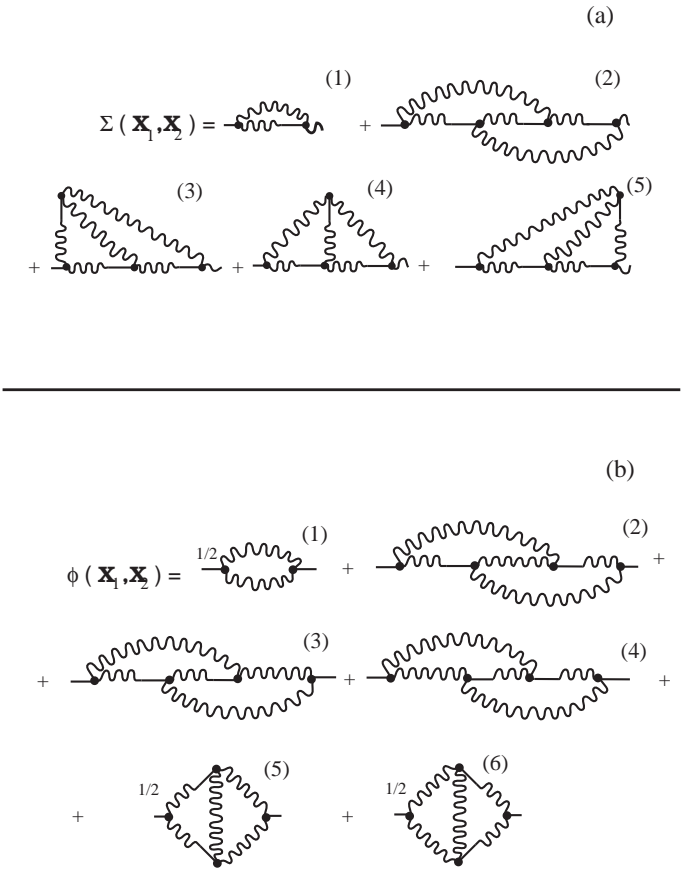}
 \vspace{.5cm}
 \caption{Diagrammatic representation of the  renormalized series
 expansion for the mass operators. (a) The mass   operator {$\Sigma$}
 of the Dyson  equation (2.8)  and (b) the mass operator {$\Phi$}
 of the Wyld equation (2.9).   }
 \label{fig:fig2}
 \end{figure}

 The starting point of the analysis are the Navier-Stokes equations for the
 velocity field of an incompressible fluid with kinematic viscosity   $\nu$
 which is forced by an external force  ${\bf f}( {\bf r},t)$:
 \def\OP{\raisebox{.2ex}{$\stackrel{\leftrightarrow}{\bf P}$}}   
 \begin{equation}
 \left( { \partial  \over  \partial  t} - \nu\Delta^{2}\right) {\bf u}
 + \OP ( {\bf u} \cdot {\bbox{\nabla}}) {\bf u } =
 \OP {\bf  f}
 \label{b1}
 \end{equation}
 where  $\OP$ is the transverse projection operator $\OP \equiv -
 \Delta^{-2} {\bbox{\nabla}}\times {\bbox{\nabla}}\times $. It was
 explained in  Paper I that developing a perturbative approach
 \cite{95LP-a,61Wyl,73MSR,78DP} for the correlation functions and response
 functions in terms of the Eulerian velocity ${\bf u}( {\bf r},t)$ results
 in a theory that is plagued with infra-red divergences. On the other hand
 one can transform to new variables, and after the transformation (which
 amounts to infinite partial resummations in the perturbation theory) one
 find a renormalized perturbation theory that is finite, without any
 divergences in any order of the expansion (cf.\cite{87BL} and I) . The new
 variables are obtained from the Belinicher-L'vov
 transformation\cite{87BL},
 \begin{equation}
 {\bf v} [ {\bf r} _{0} | {\bf r},t] \equiv {\bf u}\big[ {\bf r}+
 {\bbox{\rho}}({\bf r}_{0},t),t)\big]\,,
 \label{b2}
 \end{equation}
 where  ${\bbox{\rho}}( {\bf r}_{0},t)$ is the Lagrangian trajectory of
 a fluid point started at point ${\bf r}= {\bf r}_0$ at time $t=t_0$
 \begin{equation}
 {\bbox{\rho}}( {\bf r}_{0},t) =\int_0^t {\bf u}\big[ {\bf r}+
 {\bbox{\rho}}({\bf r}_{0},\tau),\tau\big] d\tau \ .
 \label{b3}
 \end{equation}
 The natural variables for a divergence free theory are the velocity
 {\it differences }
 \begin{equation}
 {\bf w}( {\bf r}_{0}| {\bf r},t)\equiv {\bf v}[ {\bf r}_{0}
 | {\bf r},t]- {\bf v}[ {\bf r}_{0} | {\bf r}_{0},t]\ .
 \label{b4}
 \end{equation}
 Since the averages of quantities that depend on one time only can be
 computed at $t=0$, it is clear that the average moments of these variables
 are the structure functions of the Eulerian field:
 \begin{equation}
 S_{n}(|{\bf r}- {\bf r}_{0} |  ) = \langle | {\bf w}( {\bf r}_{0}
 | {\bf r},t)  |^{n}\rangle\  .
 \label{b5}
 \end{equation}
 It was shown \cite{87BL}  that these variables satisfy the Navier
 Stokes equations, and that one can develop (cf.~I) a perturbation
 theory of the diagrammatic type in which the natural quantities
 are the Green's function $G_{\alpha\beta} ({\bf r}_{0}  |{\bf r},
 {\bf r}',t,t')$ and the correlation function $F_{\alpha\beta}
 ({\bf r}_{0} |{\bf r},{\bf r}',t,t')$:
 \begin{eqnarray}
 G _{\alpha\beta}( {\bf r}_{0}| {\bf r}, {\bf r}',t,t') &=&
 { \delta \langle w _{\alpha}( {\bf r}_{0} | {\bf r},t) \rangle
 \over  \delta h_{\beta}( {\bf r}',t')}{\Bigg |}_{h \to 0}
 \label{b6} \\
 F_{\alpha\beta}( {\bf r}_{0}| {\bf r}, {\bf r}',t,t')
 &=&
 \langle w_{\alpha}( {\bf r}_{0}| {\bf r},t) w_{\beta}
 ( {\bf r}_{0} | {\bf r}',t')\rangle\ .
 \label{b7}
 \end{eqnarray}
 In stationary turbulence these quantities depend on $t'-t$ only, and we
 can denote this time difference as  $t$. The quantities satisfy the well
 known and exact Dyson and Wyld coupled equations. The Dyson equation reads
 \begin{eqnarray}
 {\partial \over  \partial  t}&-&\nu \Delta G_{\alpha\beta}
 ( {\bf r}_{0}|{\bf r}, {\bf r}',t) = G^{0}_{\alpha\beta}
 ( {\bf r}_{0} | {\bf r}, {\bf r}',0^{+}) \delta (t)
 \nonumber \\
 &+&\int d {\bf r}_{2} G^{0}_{\alpha\delta}
 ({\bf r}_{0} | {\bf r}, {\bf r}_{2},0^{+}) \int d {\bf r}_{1}
 \int_ 0^t dt_{1}
  \label{b8}\\
 &\times&
 \Sigma_{\delta\gamma}( {\bf r}_{0}| {\bf r}_{2}, {\bf r}_{1} ,t_{1})
 G _{\gamma\beta}({\bf r}_{0}| {\bf r} _{1}, {\bf r}',t-t_{1})\,,
  \nonumber
 \end{eqnarray}
 where $ G^{0}_{\alpha\beta}( {\bf r}_{0} | {\bf r}, {\bf r}',0^{+})$ is
 the bare Green's function determined by Eq.(3.20) in paper I. We will only
 need the asymptotic properties of this function, given in
 Eqs.(\ref{b17},\ref{b18}).  The Wyld equation has the form
 \begin{eqnarray}
 &&F_{\alpha\beta}( {\bf r}_{0} | {\bf r}, {\bf r}',t)
 = \int d {\bf r}_{1} d {\bf r}_{2}
 \int_0^ \infty dt_{1} dt _{2} G_{\alpha\gamma}
 ( {\bf r}_{0} | {\bf r}, {\bf r}_1,t_{1})
 \nonumber\\
 &\times& \Big[ D_{\gamma\delta} ( {\bf r}_{1} - {\bf
 r}_{2},t-t_{1}+t_{2} ) + \Phi_{\gamma\delta}( {\bf r}_{0}| {\bf r}_{1},
 {\bf r}_{2}, t-t_{1}+t_{2}){\Big  ] }
 \nonumber\\
 &\times& G_{\delta\beta}( {\bf r}_{0} | {\bf r}', {\bf r} _{2} ,t_{2})\ .
  \label{b9}
 \end{eqnarray}
 In equation (\ref{b8}) the ``mass operator" $\Sigma$ is related to the
 ``eddy viscosity" whereas in Eq.(\ref{b9}) the ``mass operator" $\Phi$ is
 the renormalized ``nonlinear" noise which arises due to turbulent
 excitations. Both these quantities are dependent on the Green's function
 and the correlator, and thus the equations are coupled. The diagrammatic
 notations notations for the Green's function, double velocity correlator
 and vertex are presented in Fig.\ref{fig:fig1}. The diagrammatic
 representation of $\Sigma$ and $\Phi$ is well known, and is reproduced in
 Fig.\ref{fig:fig2}.  In deriving equations (\ref{b8}) and (\ref{b9}) one
 assumes that the driving force of the equation for ${\bf w}( {\bf r}_{0} |
  {\bf r},t)$ is a Gaussian random force whose covariance is the factor $D
 { _{\gamma\delta} }$ which appears in (\ref{b9}). For future reference we
 need to remind the reader of the existence of a so called ``principal
 cross section" in the diagrams for $\Phi$. This is defined as a section
 of the diagram that divides left and right parts of the diagrams
 by cutting through wavy lines of double correlators only. Each diagram in
 the series for $\Phi$ has a unique principal cross section, and see paper
 I for more details.

  The main result of I is a proof of the property of ``locality" in the
 Dyson and Wyld equations. This property means that given a value of $|{\bf
 r}- {\bf r}_{0}|$   in Eq.(\ref{b8}), the important contribution to the
 integral on the RHS comes from that region where $| {\bf r}_{1}- {\bf
 r}_{0} |$   and $| {\bf r}_{2}- {\bf r}_{0}|$  are of the order of $| {\bf
 r}- {\bf r} _{0}| $ . Moreover, the diagrams in the expansion for
 $\Sigma$, see Fig.\ref{fig:fig2}, are also local in the same sense: all
 the intermediate coordinates in all the integrals must be of the same
 order of magnitude, i.e.  $ | {\bf r}- {\bf r} _{0} | $,  in order to give
 an appreciable contribution.  In other words, all the integrals converge
 both in the upper and the lower limits. The same is true for the Wyld
 equation, meaning that in the limit of large $L$ and small  $\eta$ these
 length scales disappear from the theory, and there is no natural cutoff in
 the integrals in the perturbative theory. In this case one cannot form a
 dimensionless parameter like $L/r$ or $r/\eta$ to carry dimensionless
 corrections to the K41 scaling exponents. For $ \eta \ll | {\bf r}- {\bf
 r} _{0}| \ll L$ scale invariance prevails, and one finds precisely the K41
 scaling exponents:
 \begin{eqnarray}
 G _{\alpha\beta} (\lambda {\bf r}_{0} |  \lambda {\bf r},\lambda {\bf
 r}',\lambda ^{z} t) &=& \lambda  ^{\xi_{2}} G _{\alpha\beta}
 ( {\bf r} _{0}| {\bf r}, {\bf r}',t),
 \label{b10} \\
 F_{\alpha\beta}(\lambda {\bf r} _{0} |  \lambda {\bf r},\lambda {\bf
 r}',\lambda {     ^{z}  }t) &=& \lambda ^{\zeta_{2}}
 F_{\alpha\beta}( {\bf r}_{0}| {\bf r}, {\bf r}',t)\ .
 \nonumber
 \end{eqnarray}
 One can derive two scaling relations which hold order by order, i.e
 \begin{equation}
 2z + \zeta {     _{2 }  }=2\,,\quad  z + 2\zeta {     _{2 }  }=2 \ .
 \label{b11}
 \end{equation}
 The solution is $z = \zeta {     _{2 }  }= 2/3$. It was also shown
 that $\xi { _{2}  } = -3$.

 Note that nonperturbative effects may change these scaling relations. In
 this paper we will see that even if we use the K41 values of these scaling
 exponents, we discover anomalous scaling in turbulence for higher order
 objects. We need to look for them using nonperturbative methods. The aim
 of this paper is to present a careful study of the {\it nonlinear} Green's
 function in which anomalous exponents appear in a most transparent way.

  Some additional results that were proved in I and that we need for our
 analysis below have to do with the time integrals and time scales of the
 Green's functions in the asymptotic regimes $r  \gg  r'$ and $r  \ll
 r'$. We find that the time integral of the Green's function satisfies
 \begin{equation}
 \int_ 0^\infty dt~G(0  | {\bf r}, {\bf r}',t)) = \tau(r,r') G {     ^{0}
 }(0 | {\bf r}, {\bf r}',0 {     ^{+}  })
 \label{b12}
 \end{equation}
 where
 \begin{equation}
 \tau(r,r')   \sim \tau(r) (r'/r)^{\alpha}\quad {\rm for}\quad
    r \gg r'
 \label{b13}
 \end{equation}
 with  some scaling exponent $\alpha \geq 0$, and with
 \begin{equation}
 \tau(r) \simeq [{\bar\varepsilon}]{     ^{-1/3}  }r{     ^{2/3 }  }\ .
 \label{b14}
 \end{equation}
 In the standard phenomenology of turbulence the time scale $\tau(r)$ is
 known as the characteristic turn-over time of an $r$-eddy.  We showed that
 this time scale appears naturally in a scaling relation, i.e.
 \begin{equation}
 \tau(r)=r/\sqrt{S     _{2}  (r)}\  .
 \label{b15}
 \end{equation}
 In the opposite limit, i.e. for $r \ll   r'$, we found that
 \begin{equation}
 \tau(r,r')   \sim  \tau(r) (r'/r)^\beta
 \quad{\rm for}\quad    r\ll   r'
 \label{b16}
 \end{equation}
 where $\beta \leq  0$. The zero-time Green's function $G {     ^{0}  }(0
 | {\bf r}, {\bf r}',0 {  ^{+}  })$ which appears in (\ref{b12}) and
 serves as an initial condition for the Dyson equation has the following
 asymptotic properties:
 \begin{eqnarray}
 G^{0}_{\alpha\beta} (0  | {\bf r}, {\bf r}',0 {
 ^{+}  })   &  \sim &  (1/r')^3 \quad{\rm for}\quad  r\gg r'\,,
 \label{b17} \\
 G^{0}_{\alpha\beta}(0  | {\bf r}, {\bf r}',0 {
 ^{+}  })    &\sim &    {\bf r} { \cdot  } {\bf r}'/r' {^{5}  }
  \quad{\rm for}\quad r' \gg   r \ .
 \label{b18}
 \end{eqnarray}
 Finally, we need to quote also the results for the asymptotic properties
 of the correlator which were derived in I. For $r   \ll   r'$
 \begin{equation}
 F {     _{\alpha\beta}  } (0  | {\bf r}, {\bf r}',0)
 \sim  \bar\varepsilon{     ^{2/3 }  }\big( r {     ^{2/3 }  }+
   r/r' {     ^{1/3}  }\big) \  .
 \label{b19}
 \end{equation}
 The correlator is symmetric in $r$ and $r'$ and therefore in the opposite
 limit one just replaces $r$ and $r'$ in (\ref{b19}).
 \section{ The Nonlinear Green's Function}
 \label{19-sect3}
  In this section we begin to expose a mechanism for anomalous scaling
 which operates in the context of various many-point objects which appear
 in the theory. Such objects depend on two or more space coordinates. For
 pedagogical purposes it is convenient to discuss one of the simplest
 objects which display anomalous behavior, which is the nonlinear Green's
 function  $G_{2}( {\bf r}_{0} | x_{1} , x_{2},x_{3},x_{4})$ defined as,
 \begin{eqnarray}
 G_{2}^{\alpha\beta\gamma\delta}( {\bf r} {   _{0} } | x {
 _{1} },x {   _{2} },x {   _{3} },x {   _{4} }) &=&
 \left\langle  {\delta w {
 _{\alpha} }( {\bf r} { _{0} } | x {   _{1} }) \over \delta h {   _{\beta}
 }( {\bf r} { _{0} } | x {   _{3} })}{\delta w {   _{\gamma} }( {\bf r} {
 _{0} } | x { _{2}) }\over \delta h {   _{\delta} }( {\bf r} {   _{0} } | x
 { _{4} }) }\right\rangle\,,
 \nonumber\\
 &&\label{c1}
 \end{eqnarray}
 where for brevity we use the notation $x_{j}  \equiv   \{ {\bf r} { _{j}
 },t {   _{j} }\} $. In a Gaussian theory (which our is not) this quantity
 would be the product of the linear Green's functions, $G { ^{\alpha\beta}
 }( {\bf r} {   _{0} } | x {   _{1} },x {   _{3} })G { ^{\gamma\delta} }(
 {\bf r} { _{0} } | x {   _{2} },x {   _{4} })$. In a non-Gaussian theory
 it is natural to assume that this quantity is a homogeneous function of
 its arguments when they are in the scaling regime.  This means that
 \begin{eqnarray}
 &&G_{2}^{\alpha\beta\gamma\delta}( {\bf r} {   _{0} } |
 {\lambda}{\bf r} {   _{1} },{\lambda}{   ^{z} }t {   _{1} },{\lambda}{\bf
 r} {   _{2} },{\lambda}{   ^{z} }t {   _{2} },{\lambda}{\bf r} {   _{3}
 },{\lambda}{   ^{z} }t {   _{3} },{\lambda}{\bf r} { _{4} },{\lambda}{
 ^{z} }t {   _{4} })
 \nonumber \\
 &=& {\lambda}^{\xi_{4} }G_{2}^{\alpha\beta\gamma\delta}
  ( {\bf r} { _{0} } | x { _{1} },x { _{2} },x { _{3} },x { _{4} })
  \label{c2}
  \end{eqnarray}

 From the Gaussian decomposition of this quantity we would guess that $\xi
 { _{4} } = 2\xi {   _{2} } = -6$. The proof of locality in I means that
 there is no perturbative mechanism to change this scaling
 index. On the other hand, this quantity, which is a function of four
 space-time coordinates $x {   _{i} }$ has scaling properties that are not
 exhausted by the overall scaling exponent $\xi {   _{4} }$. We will show
 that when we consider its dependence on {\it ratios} of space-time
 coordinates in their asymptotic regimes we pick up a set of anomalous
 scaling exponents.  Our first objective is to show that in the regime $r_1
  \sim  r _2 \ll  r {   _{3 } } \sim  r {   _{4 } }$ the diagrammatic
 expansion of this object produces logarithms like ln$(r {   _{3} }/r {
 _{1} })$ to some power. Next we will prove that the sum of such
 logarithmically large contributions is given by $(r {   _{3} }/r {   _{1}
 }) {   ^{\Delta} }$ with some anomalous exponent $\Delta$. To make the
 appearance of anomalous exponents evident we begin with the simplest
 object that resums to logarithms, i.e.  the series of ``ladder diagrams".

 \subsection{The ladder diagrams }
\begin{figure}
 \epsfxsize=8.6truecm
 \vspace{.4cm}
 \epsfbox{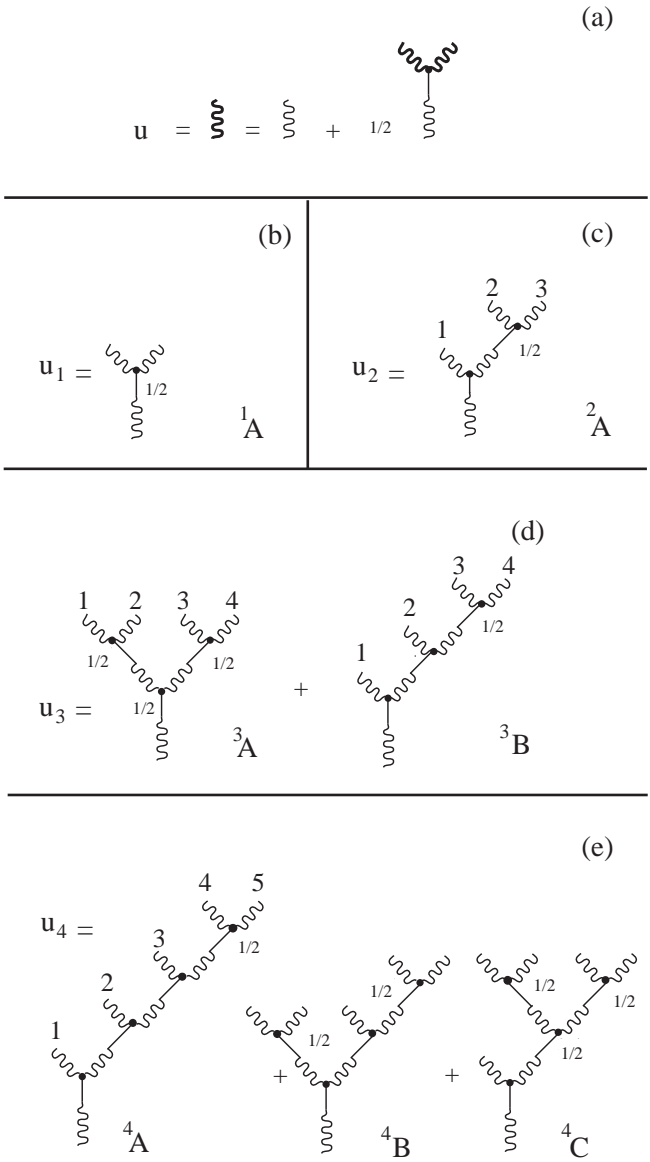}
 \caption{The diagrammatic representation of the B-L velocity. Panel (a)
 shows the equation for the renormalized velocity, and panels (b)-(e) show
 the expansion up to fourth order in the vertex.}
 \label{fig:fig3}
 \end{figure}

\begin{figure}
 \epsfxsize=8.6truecm
 \epsfbox{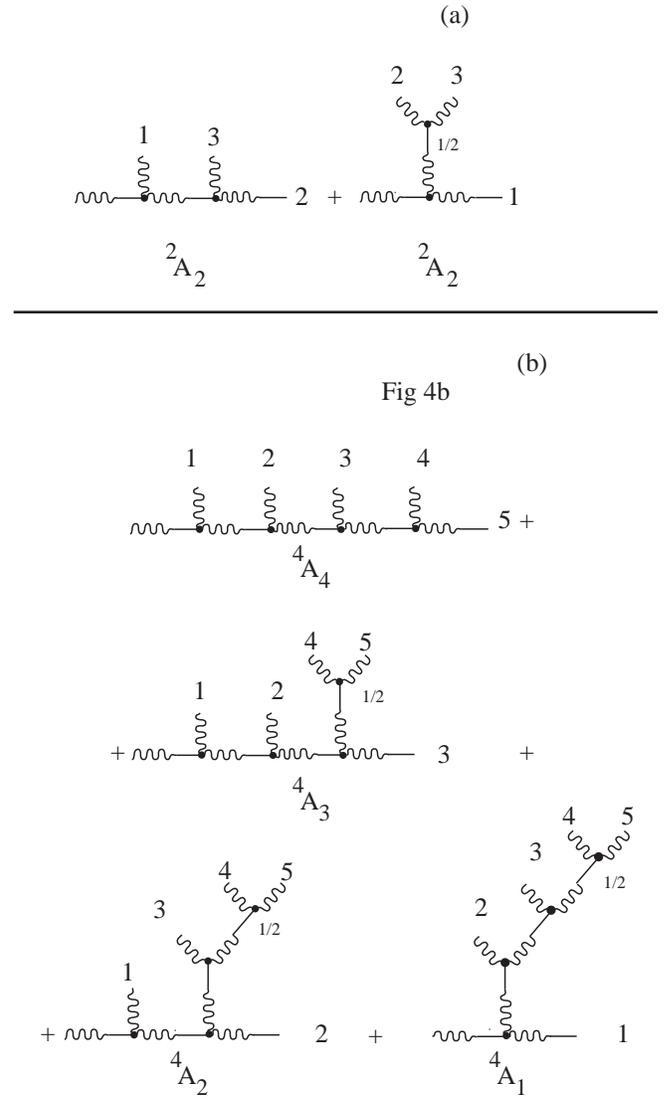}
 \vspace{.4cm}
 \caption{The diagrams for the unaveraged response, up to fourth order in
                 the vertex.}
 \label{fig:fig4} \end{figure}

\begin{figure}
 \epsfxsize=8.6truecm
 \epsfbox{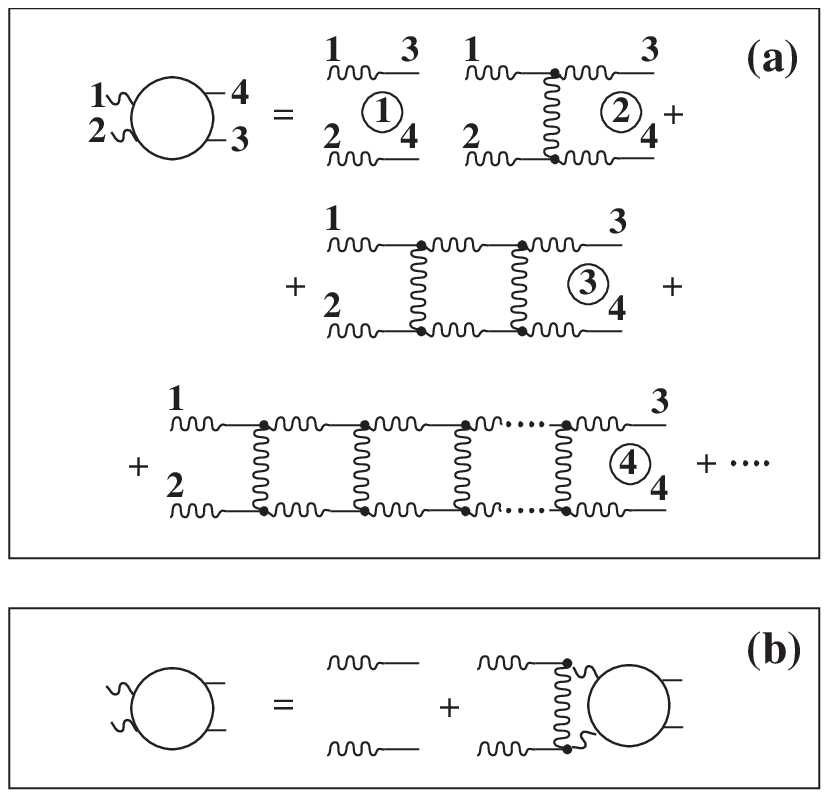}
 \vspace{.5cm}
 \caption{The diagrammatic expansion of the 4-point nonlinear Green's
         function (3.1). In this figure we show the simple ladder diagrams
         in panel (a) and the resummation of the simple ladder diagrams in
         panel (b)}
 \label{fig:fig5} \end{figure}

 The diagrammatic representation of the nonlinear Green's function
 (\ref{c1}) is obtained as follows. In the spirit of the Wyld
 expansion\cite{61Wyl,95LP-a}  one can begin with the diagrams for ${\bf
 w}$ which are shown in Fig.\ref{fig:fig3}.
 After differentiating with respect to a force we get the diagrams for the
 unaveraged response ${\delta w} { _{\alpha} }( {\bf r} {   _{0} } | x {
 _{1} })/{\delta h} {   _{\beta} }( {\bf r} { _{0} } | x { _{3} })$ shown
 in Fig.\ref{fig:fig4}.  We recall (see I) that each of these diagrams has
 a principal path of Green's functions connecting an entry denoted by a
 wavy line to an exit denoted by a straight line. At this point we can take
 any combination of two diagrams, one for ${\delta w} { _{\alpha} }( {\bf
 r} { _{0} } | x {   _{1} })/{\delta h} {   _{\beta} }( {\bf r} { _{0} } |
 x {   _{3} })$ and one for ${\delta w} { _{\alpha} }( {\bf r} { _{0} } | x
 {   _{2} })/{\delta h} { _{\beta} }( {\bf r} { _{0} } | x {   _{4} })$,
 glue them together according to the Gaussian rules and then perform the
 Dyson-Wyld line resummation. Clearly every resulting diagram has two
 principal paths, one going from $x {   _{1} }$ to $x { _{3} }$ and the
 other from $x {   _{2} }$ to $x {   _{4} }$, see Fig.\ref{fig:fig5}a.
 Every diagram begins with two wavy lines and ends with two straight lines.
 One infinite sum of contributions results from having all the averaging
 done in each tree separately. Such a sum results in the first diagram on
 the RHS of Fig. 4a, which is precisely the product of two {\it dressed}
 Green's functions.  Next we have infinite number of diagrams in which
 these two principal paths are connected via fragments that resum to a
 dressed correlator. This is the sum that gives rise to the second diagram
 on the RHS of Fig.  4a.  Following are infinite sums that result in
 connection via two, three etc.  dressed correlators.  The sum of terms
 appearing in Fig.\ref{fig:fig5}a is referred to as the sum of ``simple
 ladder diagrams".  We note that this sum can be represented exactly in
 graphical notation as shown in Fig.\ref{fig:fig5}b. We used here
 explicitly the simple topology of the simple ladder diagrams.

\begin{figure}
 \epsfxsize=8.6truecm
 \epsfbox{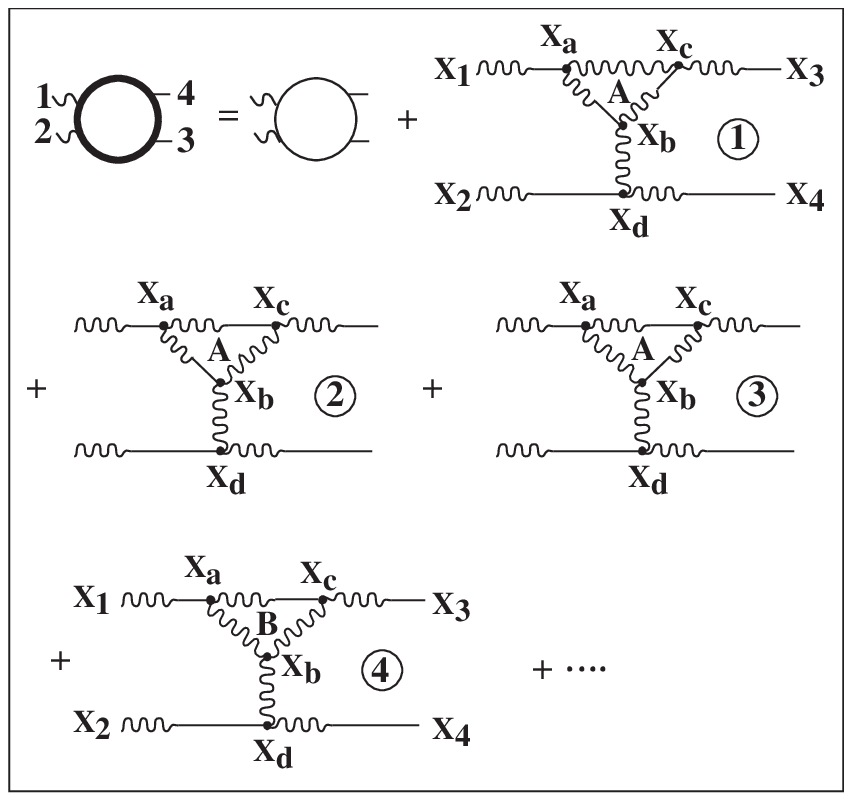}
 \vspace{.5cm}
 \caption{The diagrammatic expansion of the 4-point nonlinear Green's
 function (3.1).  In this figure we show diagrams that add to those shown
 in Fig.5, and that decorate the vertices. The sum of these diagrams turns
 the vertices into dressed ones.}
 \label{fig:fig6}
 \end{figure}

  In the simpler case of the passive scalar that was presented in\cite{94LPF}
 the simple ladder diagram tell the whole story. This is not the case here.
 There exist an infinite class of diagrams that decorate the simple ladder
 diagrams as shown in Fig.\ref{fig:fig6}.  The result of the summation
 of all the simple ladder diagrams in Fig.\ref{fig:fig5} which is denoted
 as a thin circle is added to the diagrams that contain decorations of the
 vertices, see Fig.\ref{fig:fig6}. These decorations are identified as
 contributions that can be resummed to dress the vertices. In these
 diagrams there are two types of dressed vertices that we denote as A and B
 respectively. The decorated vertex A has one straight tail and two wiggly
 tails like the bare vertex, and therefore A can be considered as the
 dressing of the bare vertex. The vertex B has two straight tails and one
 wiggly, and it does not have a bare counterpart.

\begin{figure}
 \epsfxsize=8.6truecm
 \epsfbox{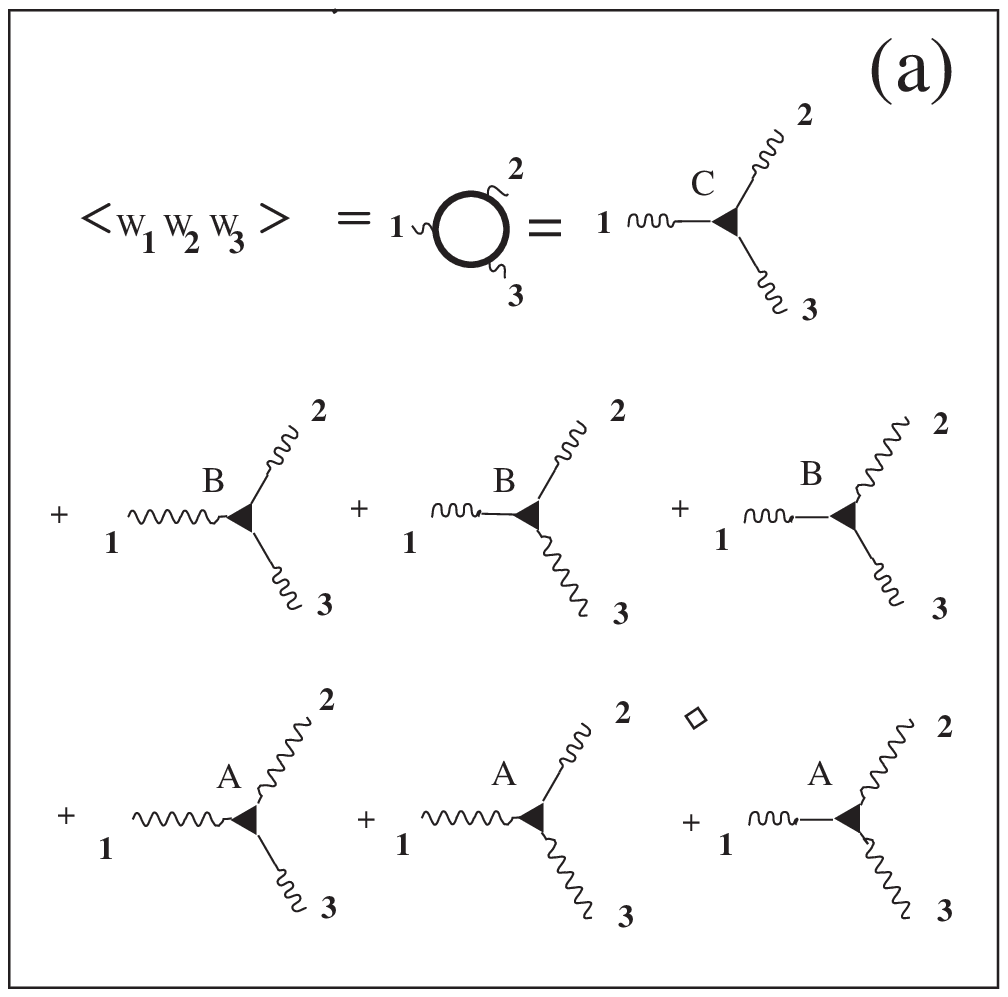}
 \epsfbox{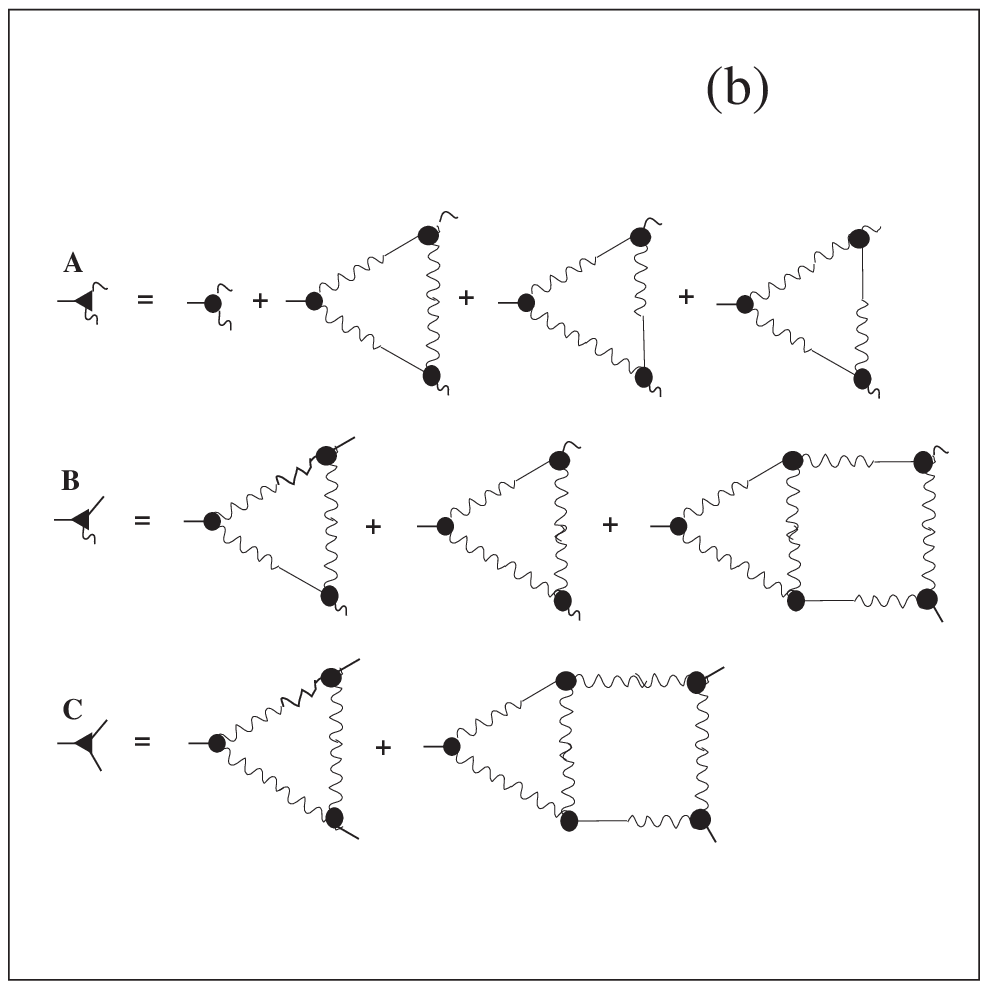}
 \caption{The diagrams that dress the vertices. In this paper only vertices
 of type A and B appear. }
 \label{fig:fig7}
 \end{figure}

\begin{figure}
 \epsfxsize=8.6truecm
 \epsfbox{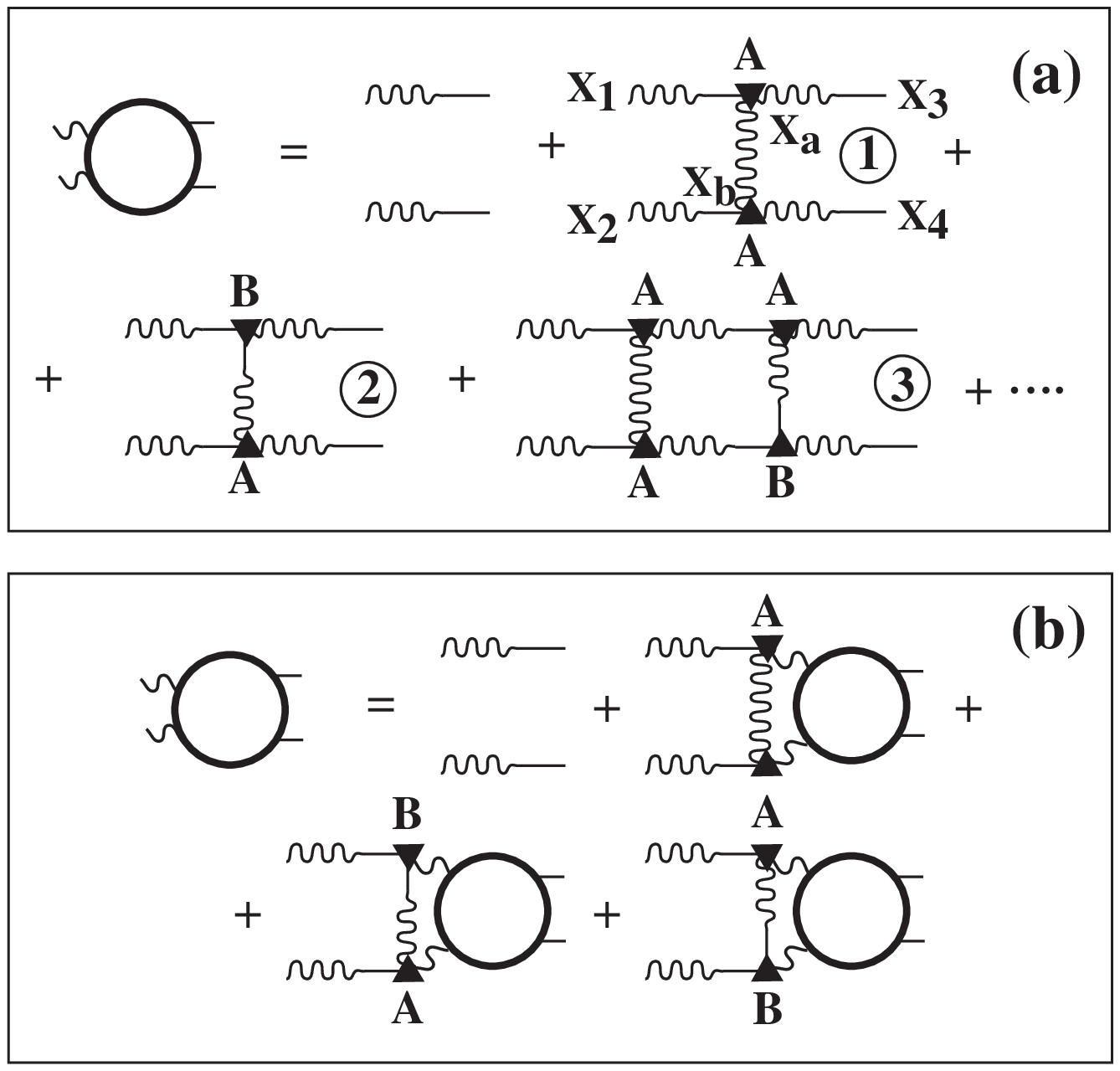}
 \vspace{.3cm}
 \caption{The diagrammatic expansion of the 4-point
         nonlinear Green's function (3.1) In this figure we show the
         dressed simple ladder diagrams, panel (a) and the resummed
         equation (panel (b) which results from summing all the
         diagrams in panel (a) .}
  \label{fig:fig8}
 \end{figure}

 In the general theory one finds an infinite series of diagrams that dress
 the vertices A and B, and this series is shown in Fig.\ref{fig:fig7}. It
 is therefore clear that the sum of all the diagrams appearing in
 Fig.\ref{fig:fig5} and in Fig.\ref{fig:fig6} can be represented as
 diagrams that have the topology of the simple ladder diagrams but with
 dressed vertices. These diagrams are shown in Fig.\ref{fig:fig8}. We refer
 to the diagrams shown in panel a of Fig.\ref{fig:fig8} as the sum of
 ``dressed simple ladder diagrams", and denote the result of this summation
 with a bold circle. We note that one cannot have two vertices of type B in
 one rung of the ladder. Such a rung will call for a connection via a long
 straight line, an object that does not exist in this theory.  Panel b of
 Fig.\ref{fig:fig8} represents the exact resummation of all the diagrams
 having the topology of the dressed simple ladder. Note that the resummed
 series which is represented as a heavy circle appears on both the LHS and
 RHS of this equation. This is not yet the fully resummed nonlinear Green's
 function since there are additional diagrams with different topology.

\begin{figure}
 \epsfxsize=8.6truecm
 \epsfbox{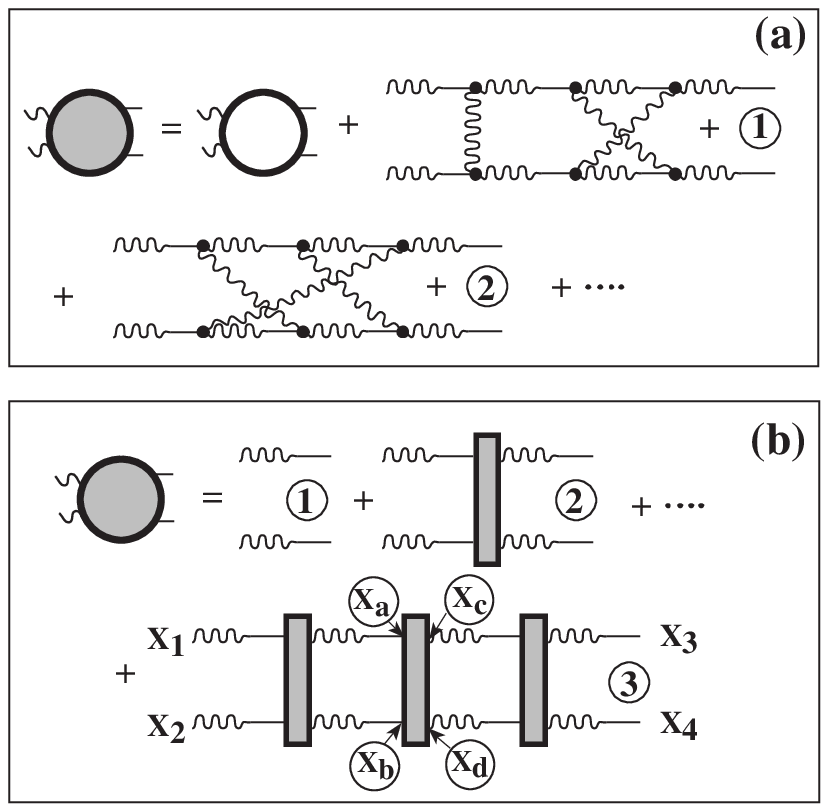}
 \vspace{.5cm}
 \caption{The diagrammatic expansion of the 4-point
         nonlinear Green's function (3.1) . Here we show the fully
          renormalized quantity, which is obtained by summing the
         result of the summation in Fig.8 to all the more
         complex diagrams that appear in the series. The
         dark rungs of the ladder are the two-eddy
         irreducible mass operator, whose diagrammatic
         expansion is shown in Fig.10.}
 \label{fig:fig9}
 \end{figure}

 Next we need to discuss the diagrams whose topology differs from that of
 the dressed simple ladder diagrams. Such diagrams are obtained when the
 connections between the principal paths intersect, or when there appear
 decorations that cannot be absorbed into the dressed vertices. The
 simplest examples are shown in Fig. \ref{fig:fig9}a, in which the vertices
 have been already dressed by summing the corresponding decorations of the
 diagrams shown.

 The resummation of all the diagrams for the Nonlinear Green's function is
 shown in Fig.\ref{fig:fig9}b, where the series for the renormalized rung
 is given in Fig. 9a. The diagrams in the series can be classified into two
 classes, which we refer to as ``two-eddy reducible" and ``two eddy
 irreducible" diagrams. To define these we need first to define ``interior"
 Green's functions as Green's functions belonging to one of the two
 principal paths but not to an entry or an exit. ``Two-eddy reducible"
 diagrams are diagrams that can be cut into two pieces by cutting through
 two interior Green's functions belonging to two principal paths without
 cutting any other line. ``Two-eddy irreducible" diagrams are diagrams that
 are not two-eddy reducible. As examples consider the diagrams in
 Fig.\ref{fig:fig5}. In 4a the two first diagrams are two-eddy irreducible,
 and the next two are two-eddy reducible. In Fig.\ref{fig:fig9} diagram 1
 is two-eddy reducible whereas diagram 2 is two-eddy irreducible.

\begin{figure}
 \epsfxsize=8.6truecm
 \epsfbox{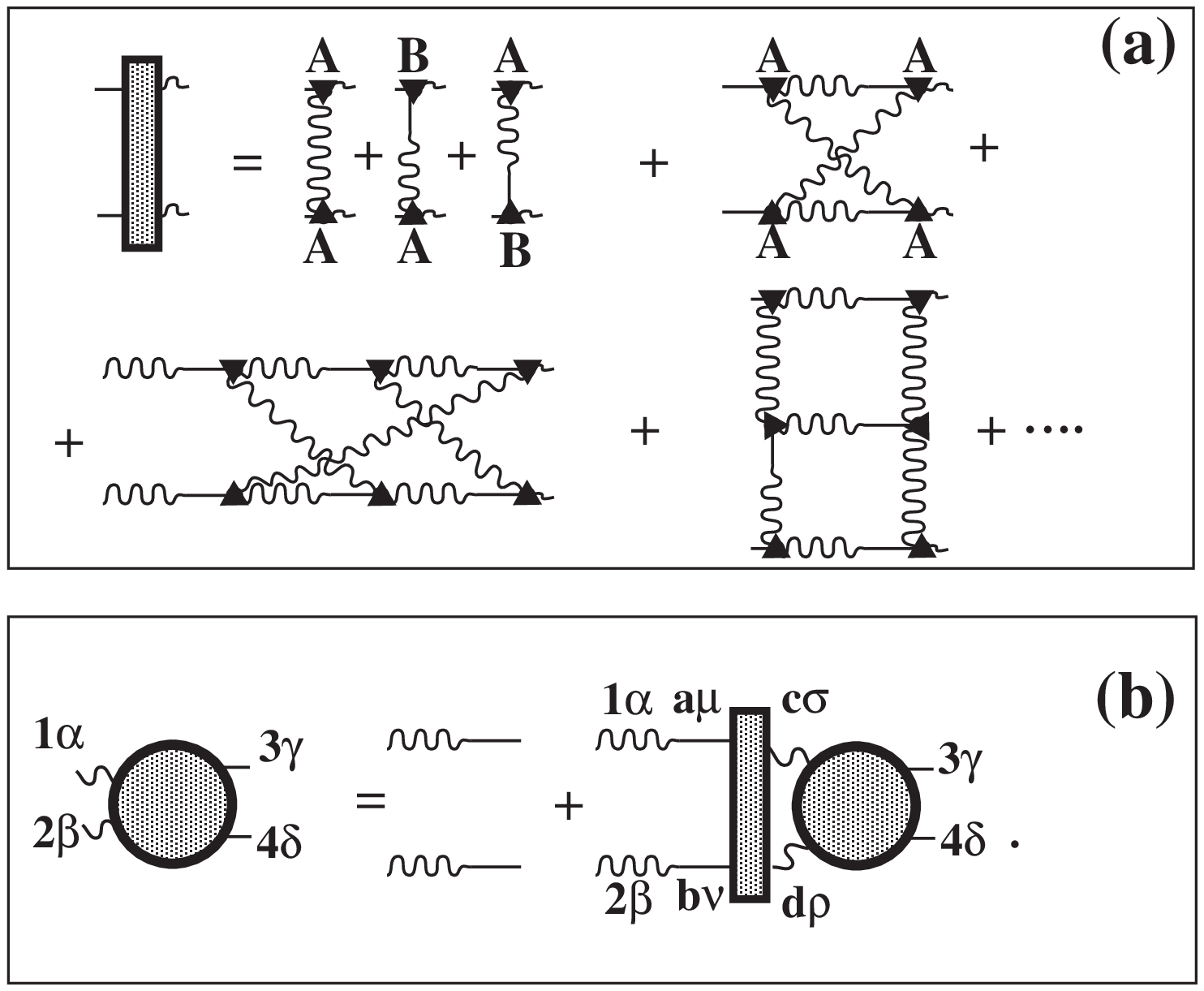}
 \vspace{.5cm}
 \caption{(a) the diagrammatic series for the
         two-eddy irreducible mass operator. (b) The integral
         equation for the fully renormalized nonlinear Green's
         function.}
 \label{fig:fig10}
 \end{figure}

 The sum of all the two-eddy irreducible diagrams gives rise to the ``two
 eddy mass operator" ${\Sigma}{   _{2} }( {\bf r} {   _{0} } | x {   _{1}
 },x { _{2} },x {   _{3} },x {   _{4} })$. This object is given in terms of
 the infinite expansion shown in Fig.\ref{fig:fig10}a. Using this object we
 can sum all the diagrams belonging to the nonlinear Green's function, as
 shown in Fig.\ref{fig:fig10}b.  This is an exact equation for the
 nonlinear Green's function, which is the direct analog of the Dyson
 equation for the usual Green's function. Note that the fully resummed
 nonlinear Green's function is denoted by a gray circle with bold
 circumference. Thus in Fig. 8a the bold empty circle which is the sum of
 all the dressed simple ladder diagrams is added to all the diagrams that
 have more complex topology, to yield the gray circle object. Note that the
 equation in Fig.\ref{fig:fig10}b reduces to the equation in
 Fig.\ref{fig:fig8}b if we substitute the first three terms of
 Fig.\ref{fig:fig10}a instead of the two eddy mass operator in
 Fig.\ref{fig:fig10}b.

 \subsection{Logarithmic contributions in simple ladder diagrams}
  In this sub-section we demonstrate that the leading contribution to the
 ladder diagrams involves logarithmically large
 terms\cite{95LP-a,95LP-b,95LP,95LL,94LL,94LPF}.  To this aim we introduce
 the time integral of the nonlinear Green's function defined by
 \begin{eqnarray}
 &&{\bf G}_{2}^{tt}( {\bf r} {   _{0} } | x {   _{1} },x { _{2} }, {\bf r}
 { _{3} }, {\bf r} {   _{4} }) \equiv \int dt {   _{3} }dt {   _{4 } } {\bf
 G} {   _{2} }( {\bf r} {   _{0} } | x {   _{1} },x { _{2} },x { _{3} },x {
 _{4} })\ .
 \nonumber\\
 \label{c3}
 \end{eqnarray}
 The zero order contribution to this quantity is obtained from the first
 term in Fig.\ref{fig:fig5}a which is the product of two linear Green's
 functions and is denoted as diagram 1. Using Eq.(\ref{b12}) we compute
 \begin{eqnarray}
 && {\bf G}_{2,0}^{tt}( {\bf r} {   _{0} } | x {   _{1} },x {
 _{2} }, {\bf r} {   _{3} }, {\bf r} {   _{4} })
 \label{c4} \\
  &=& \tau(r {   _{1} },r { _{3} }) \tau(r {   _{2} },r {   _{4} }) {\bf G}
 {   _{0} }( {\bf r} {   _{0} } | r {   _{1} },r {   _{3} },0 {   ^{+} })
 {\bf G} {   _{0} }( {\bf r} { _{0} } | r {   _{2} },r {   _{4} },0 {
 ^{+} })\ .
 \nonumber
 \end{eqnarray}
 Consider now the regime $r {   _{1} }  \sim  r {   _{2} }  \ll  r {   _{3}
 }  \sim  r {   _{4} }$. We can  use Eq.(\ref{b13}) to estimate
 \begin{equation}
 {\bf G}_{2,0}^{tt}( {\bf r} {   _{0} } | x {   _{1} },x {
 _{2} }, {\bf r} {   _{3} }, {\bf r} {   _{4} })
 \sim  {
 \tau(r_{1})\tau(r _{2})\over (r_{3}r _{4})^{3}
 }
 \Big( {r_{1}r_{2}\over r_{3}r_{4} }\Big)
 ^{\beta +1}\ .
 \label{c5}
 \end{equation}
 Consider now the next term which is the shortest ladder in
 Fig.~\ref{fig:fig5} (i.e the ladder with one rung and two vertices which
 is labeled as diagram 2).  It is denoted as ${\bf G}_{2,2}^{tt}( {\bf r} {
 _{0} } | x {   _{1} },x {   _{2} }, {\bf r} {   _{3} }, {\bf r} {   _{4}
 })$ and it is written as
 \begin{eqnarray}
 &&
 \!\!\!\!\!\!\!\!
 {\bf G}_{2,2}^{tt}( {\bf r} {   _{0} } | x {   _{1} },x { _{2} }, {\bf
 r} {   _{3} }, {\bf r} {   _{4} })\!\!  \sim \!\!  \int dx {   _{i} }dx {
 _{j } }G( {\bf r} {   _{0} } | x {   _{1} },x {   _{j} }) {   _{ } }G(
 {\bf r} { _{0} } | x {   _{2} },x {   _{i} })
 \nonumber\\
 &\times&{\partial \over \partial r {   _{i} }}
 {\partial \over \partial r {   _{j} }}
 F({\bf r} {   _{0} } | x {   _{i} },x {   _{j} } )
 \int dt {   _{3} }dt { _{4 } }G( {\bf r} {   _{0} } | x { _{j} },x {
 _{3} }) {   _{ } }G( {\bf r} {   _{0} } | x {   _{i} },x { _{4} })\ .
 \nonumber\\
 \label{c6}
 \end{eqnarray}
 In writing Eq.(\ref{c6}) we dropped the tensor indices from ${\bf G}$ and
 ${\bf F}$. The range of integration that is of interest for us is the
 range $r {   _{1} } \sim r {   _{2 } } \ll  r {   _{i} },r {   _{j} } \ll
 r { _{3} } \sim r {   _{4} }$. In this range we can use Eq.(\ref{b13}) and
 evaluate the integral (\ref{c6}) as
 \begin{eqnarray}
 &&{\bf G}_{2,2}^{tt}
 ( {\bf r} {   _{0} } | x {   _{1} },x { _{2} }, {\bf r} {   _{3} }, {\bf
 r} {   _{4} })
 \sim {  \tau(r_{1}) \tau(r_{2}) \over (r_{3}r _{4})^{3} }
 \Big( { r_{1}r_{2} \over r_{3}r_{4} }\Big)^{\beta +1}
 \nonumber \\
 &\times& \int_ {r_{1}}^{r_{3}}
 { dr {   _{i} }\over r_{i}^{2} }
 \int _{r_{2}}^{r_{4}}  { dr_{j}\over r _{j}^{2} }
 F( {\bf r} { _{0} } | r { _{i} },r { _{j} })\tau(r { _{i} })
 \tau(r { _{j} })
 \label{c7} \\
 && {\rm (provided}\quad
  r {   _{1} } \sim r {   _{2 } } \ll  r {   _{i} },r { _{j} } \ll r {
 _{3} } \sim r {   _{4} })\ .
 \nonumber
 \end{eqnarray}
 To proceed we will use the K41 scaling exponents to evaluate the
 correlators.  In other words we evaluate $F( {\bf r} {   _{0} } | r {
 _{i} },r {   _{j} })$ as min$ \{ r {   _{i} } {   ^{2/3} },r {   _{j} } {
 ^{2/3} } \}$ according to Eq.(\ref{b19}). This is not an exact step, and
 is used to demonstrate in the swiftest way the existence of logarithmic
 divergences which lead to anomalous scaling. In Section 6 we will derive
 an exact sum rule that will establish the existence of the anomalous
 exponent and will determine its value.

 Consider now separately the two possibilities $r {   _{i} }  <  r {   _{j}
 }$ and $r {   _{i} }  >  r {   _{j} }$. In the former case we find that
 the $r {   _{i } }$ integral contributes mostly in the upper limit and the
 $r {   _{j} }$ integral in the lower limit. In the latter case the
 situation is inverted. In both cases the integrals have their largest
 contribution in the regime $r {   _{i} } \sim r {   _{j } }$.  We thus can
 evaluate (\ref{c7}) as
 \begin{eqnarray}
 &&\!\!\!\!\!
 {\bf G}_{2,2}^{tt}
 ( {\bf r} {   _{0} } | x {   _{1} },x { _{2} }, {\bf r} {   _{3} }, {\bf
 r} {   _{4} })
 \sim
 {\bf G}_{2,0}^{tt}
 ( {\bf r} {_{0} } | x {_{1} },x {_{2} }, {\bf r} {_{3} }, {\bf r} {_{4}})
 \int _{r_1}^{r_3} { dr_{i} \over r_{i} }
 \nonumber \\
 &=& {\Delta}_{0}{\bf G}_{2,0}^{tt}
 ( {\bf r}_{0}| x _{1},x _{2}, {\bf r}_{3}, {\bf r}_{4})
 \ln ( r_{3}/ r_{1})\, ,
 \label{c8}
 \end{eqnarray}
 where ${\Delta} {   _{0} }$ is a dimensionless constant of $O(1)$. The
 next order contribution is
 \begin{eqnarray}
 &&     \!\!\!\!\!\!\!
 {\bf G}_{2,4}^{tt}
 ( {\bf r} {   _{0} } | x {   _{1} },x { _{2} }, {\bf r} {   _{3} }, {\bf
 r} {   _{4} })
 \!\sim \!\!\! \int \!\! dx {   _{i} }dx {   _{j } }
 G( {\bf r} {   _{0} } | x { _{1} },x { _{j} }) G( {\bf r} {
 _{0} } | x {   _{2} },x { _{i} })
 \nonumber\\
 &\times& \!\! { \partial \over \partial r {   _{i} }}
 { \partial \over  \partial r {   _{j} }}
  F( {\bf r} {   _{0} } | x { _{i} },x {   _{j} } )
 \!\! \int \!\! dx {   _{n} }dx {   _{m } }
 G( {\bf r} {   _{0} } | x {   _{j} },x { _{n} })
 G( {\bf r} { _{0} } | x {   _{i} },x {   _{m} })
 \nonumber\\
 &\times& \!\! {\partial \over  \partial r {   _{n} }}
 {\partial \over \partial r{   _{m} }}
  F( {\bf r} {   _{0} } | x {   _{n} },x {   _{m} } )
 \!\!\! \int \! \! \! dt {   _{3} }dt {   _{4 } }
 G( {\bf r} {   _{0} } | x {   _{n} },x { _{3} }) {   _{ } }
 G( {\bf r} { _{0} } | x {   _{m} },x {   _{4} }).
 \nonumber\\
 \label{c9}
 \end{eqnarray}
 It can be seen in analogy to the situation in (\ref{c7}) that the
 intermediate integrations in (\ref{c9}) peak in the regime $r {   _{i } }
 \sim  r {   _{j} }$ and $r {   _{n } } \sim  r {   _{m} }$. The largest
 contribution to the integral then comes from the regime $r {_{1} } \sim
 r {   _{2 } } \ll  r {   _{i} } \sim r {   _{j} }  \ll  r {   _{n} } \sim
 r {   _{m} }  \ll  r {   _{3} } \sim r {   _{4} }$. In this regime the
 integral is evaluated as
 \begin{eqnarray}
 &&
 {\bf G}_{2,4}^{tt}
 ( {\bf r} {   _{0} } | x {   _{1} },x { _{2} }, {\bf r} {   _{3} }, {\bf
 r} {   _{4} })
  \nonumber \\
 &\sim &{\bf G}_{2,0}^{tt}
 ( {\bf r} {   _{0} } | x {   _{1} },x {   _{2} }, {\bf r} { _{3} }, {\bf
 r} {   _{4} })
 \int_{r _{1} }^{r _{3} }
 { dr {   _{i} }\over r {   _{i} }}
 \int_{r _{i} }^{r _{3} }
 { dr {   _{n} }\over r {   _{n} }}
 \label{c10}\\
 &=& {\bf G}_{2,0}^{tt}
 ( {\bf r} {   _{0} } | x {   _{1} },x {   _{2} }, {\bf r} {   _{3} }, {\bf
 r} {   _{4} })
 {1\over 2}\Big[ \Delta { _{0} } \ln \Big({ r { _{3} }\over r {   _{1}
 }} \Big)\Big]^{2 }\ .
 \nonumber
 \end{eqnarray}
 Note that we have asserted here without proof that the coefficient in
 front of the logarithm is the square of the coefficient in (\ref{c8}).
 This is intuitively acceptable because of the repetitive nature of the
 ladder structure. It is however an important point and therefore we will
 prove it in section 5 by analyzing the resummed equation for the nonlinear
 Green's function.

  It becomes believable now that the ladder with $n$ rungs contains a
 contribution of order $\big[ \Delta_{0} \ln ( r { _{3} }/r {   _{1} })
 \big]^{n}/n!$. We reiterate that there are additional contributions from
 other regimes of the intermediate integrations and they will contribute
 lower powers of logarithms. One can take them into account, but this only
 serves to renormalize the value of ${\Delta} { _{0} }$, as is proven
 below.  At this level of demonstration it is enough to take the leading
 order contribution from each n-rung ladder and notice that the summation
 of all these contributions will give a term proportional to $(r _{3}
 /r { _{1} })^{\Delta_ 0}$.

  Next we need to prove that the decorations of the simple ladder diagrams
 do not change the qualitative picture of anomalous scaling discussed
 above.

 \section{Proof of Rigidity}
 \label{19-sect4}
\begin{figure}
 \epsfxsize=8.6truecm
 \epsfbox{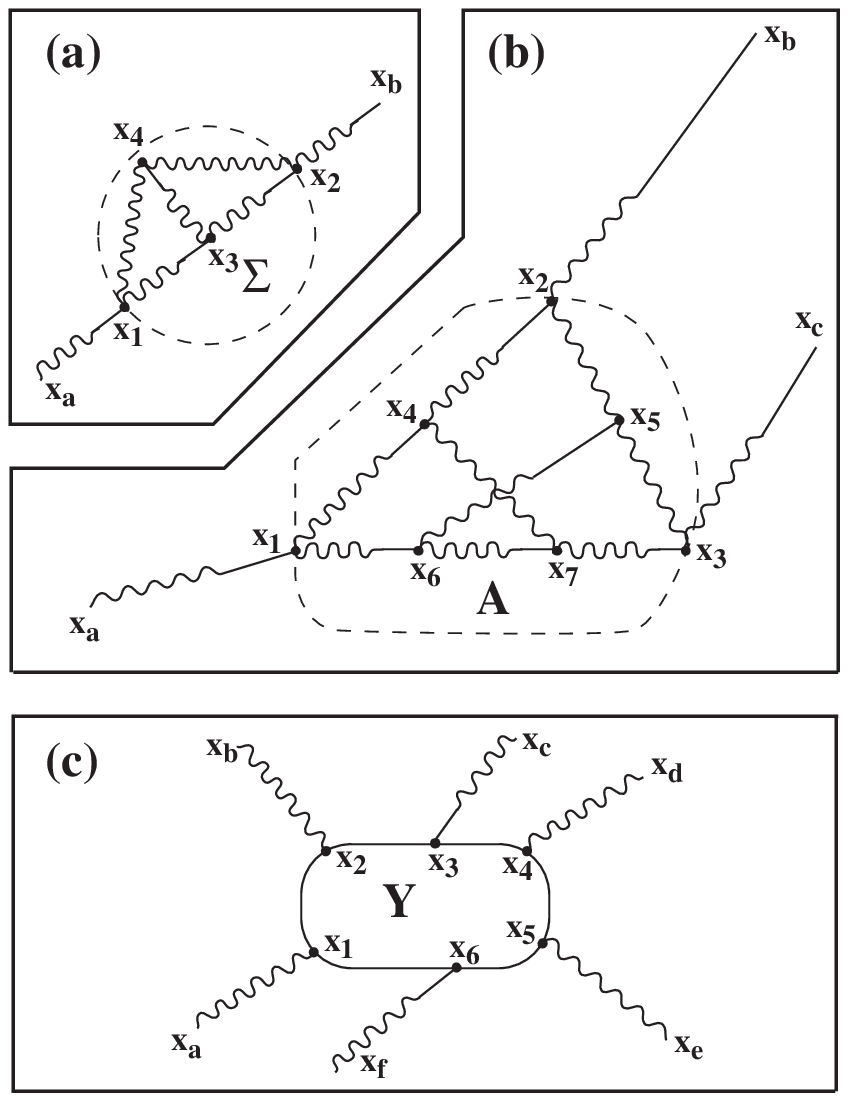}
 \vspace{.5cm}
 \caption{Diagrams used in the explanation of the property
         of rigidity. }
 \label{fig:fig11}
 \end{figure}

  A useful concept in the demonstration of anomalous scaling is the concept of
 rigidity, which is an order by order property of diagrams that we are going
 to use
 repeatedly. We explain the concept with the help of Fig.\ref{fig:fig11}.
 Consider the diagram in Fig.\ref{fig:fig11}a.  The fragment within the
 dashed circle is a part of the mass operator $\Sigma$.  Suppose that $x {
 _{a}}$ is smaller than $x { _{b}}$. the property of {\it locality} which
 was proven in I means that all the significant contributions to the
 diagrams come from the range of integration in which $x { _{1}}, x {
 _{2}}, x { _{3 }}$ and $x { _{4 }}$ are all lying between $x { _{a}}$ and
 $x { _{b}}$. The property of {\it rigidity} is stronger, and it relates to
 fragments of diagrams whose exterior points are vertices.  For the diagram
 in Fig.\ref{fig:fig11}a rigidity says that if we fix the coordinate ${\bf
 x}\- { _{1 }}$ then for $x { _{a }}\ll x { _{1}}$ and $x { _{b }}\gg x {
 _{1}}$ the largest contribution to the diagram comes from the regime of
 integration $x { _{2 }}\sim x { _{3 }}\sim x { _{4}}$ and all these
 space-time coordinates are of the order of $x { _{1}}$. We call this
 property ``rigidity" to give the intuitive feeling that one can stretch at
 will the diagram such that $|x { _{a}}-x { _{1}}|$ and $|x { _{b}}-x {
 _{1}}|$ become very large, but because $x { _{1}}$ was determined, the
 positions $x { _{2}},x { _{3}},x { _{4}}$ are rigidly fixed to the
 vicinity of $x { _{1}}$.  We call the diagram rigid if it has this
 property with regards to fixing either $x { _{1}}$ or $x { _{2}}$.  In
 diagrams that have three external legs to them, like the one in
 Fig.\ref{fig:fig11}b which belongs to the series of 3-point correlator,
 rigidity of the diagram implies that the inner fragment is rigid with
  respect to fixing either $x { _{1 }}$ or $x { _{2}}$, or $x { _{3}}$.
 Thus if we fix $x { _{1}}$ than $x { _{a}}, x { _{b}}$ and $x { _{c}}$ can
 be brought arbitrarily far away from $x { _{1}}$, and yet the main
 contribution to the integral comes from the regime $x { _{2}},\dots ,x {
 _{7}}$ are all of the order of $x { _{1}}$, etc.  This definition of
 rigidity extends to diagrams with an arbitrary number of external
 coordinates, like in Fig.  10c. Such a diagram is called rigid if fixing
 any of the coordinates $x { _{1}},...,x { _{6}}$ results in the main
 contribution to the diagram coming from the regime in which the rest of
 these coordinates are of the same order as the fixed one, independently of
 the positions of the external coordinates $x { _{a}},\dots ,x { _{f}}$.

 Lastly we want to clarify the concept of a ``stretched" diagram.  A
 stretched diagram is a a diagram in which there is a definite ordering in
 the positions of the vertices that are being integrated upon. Consider for
 example the diagram in Fig.\ref{fig:fig11}c.  Suppose that the positions
 $r { _{2}}, r { _{3 }}, r { _{4 }}$ of the group of vertices $x { _{2}}, x
 { _{3 }}, x { _{4}}$ are larger than the positions $r { _{1}}, r { _{6}},
 r { _{5}}$ of the group $x { _{1}}, x { _{6}}, x { _{5}}$. We call this
 diagram stretched if all the positions $ r { _{a}},r { _{f}}$ and $r {
 _{e}}$ are smaller than any of $r { _{1}},r { _{6}} {  }, r { _{5}}$, and
 the positions $r { _{b}}, r { _{c}}, r { _{d}}$ are all larger than $r {
 _{2}}, r { _{c}}, r { _{d}}$.  Rigidity of a fragment will be intuitively
 understood as the resistance to stretching.  In other words, one can think
 of the propagators as springs that are being pulled to stretch the
 diagram. If it is sufficient to fix one coordinate of the fragment, in
 this case any of the coordinates $x { _{1}}-x { _{6 }}$ such that the main
 contribution to the diagram comes from the regime that all the coordinates
 in this group are of the same order, we call the diagram rigid.

 We prove now that the ladder diagram are rigid.

 \subsection{Rigidity of the dressed vertices}
 The aim of this sub-section is to answer the following question: how
 should we evaluate a dressed vertex in a stretched diagram. Every dressed
 vertex has three coordinate designations, see for example
 Fig.\ref{fig:fig6}, in which these coordinates are $x { _{a}}, x { _{b}}$
 and $x { _{c}}$ respectively.  For fixed coordinates $x { _{1}}$ to $x {
 _{4 }}$ and $x { _{b }}, x { _{d }}$ we need to integrate over $ x {
 _{a}}, { _{ }}$and $x { _{c}}$. We will show here that the main
 contribution to the integral comes from the region $x { _{a}} \sim x {
 _{b}}\sim x { _{c}}$.  As a result we will be able to estimate the dressed
 vertex as $1/r { _{b}}$. Finally this estimate will allow us to repeat the
 argument that $r { _{b}}$ contributes mostly in the vicinity of $r {
 _{d}}$. Thus the estimate leading to the logarithmic divergence will
 remain unchanged.

\begin{figure}
 \epsfxsize=7truecm
 \epsfbox{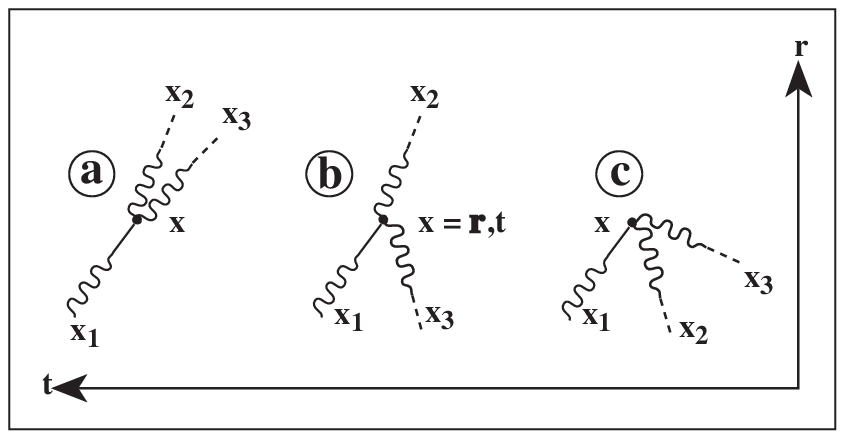}
 \vspace{.5cm}
 \caption{Some local geometries about a given vertex that appear in the
         proof of rigidity.}
 \label{fig:fig12}
 \end{figure}

\begin{figure}
 \epsfxsize=7truecm
 \epsfbox{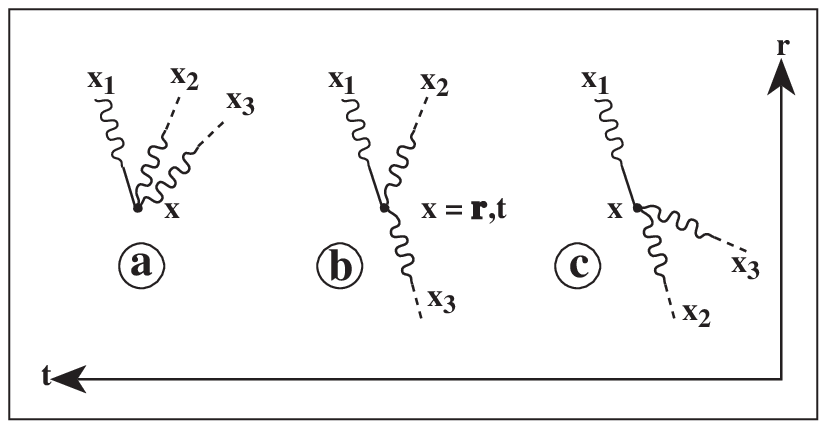}
 \vspace{.5cm}
 \caption{Some local geometries about a given vertex that appear in the
         proof of rigidity.}
 \label{fig:fig13}
 \end{figure}

 We begin by considering the various possible integrals that depend on the
 local geometry around a given vertex in a stretched diagram. These are the
 local geometries shown in Figs.\ref{fig:fig12} and \ref{fig:fig13}.  In
 these figures the dashed line represents either a straight or a wiggly
 half line. We think of these fragments as parts of a bigger diagram in a
 ``stretched" configuration, meaning that in the larger diagram there are
 no turn backs in the positions of other vertices that are not shown in
 Figs.\ref{fig:fig12} and \ref{fig:fig13}.  In other words, if we consider
 Fig.\ref{fig:fig12}a, the positions of all the vertices attached to $x {
 _{1}}$ are smaller than $r { _{1}}$, and the vertices attached to $x {
 _{2}}$ and $x { _{3}}$ are all larger than $r { _{2}}$ and $r { _{3}}$.
 For 11b it means that all the vertices attached to $x_{1}$ and $x { _{3}}$
 have positions smaller than $r { _{1}}$ and $r { _{3}}$, and all the
 vertices attached to $x_{2}$  have positions larger than $r { _{2}}$.

\begin{figure}
 \epsfxsize=8.6truecm
 \epsfbox{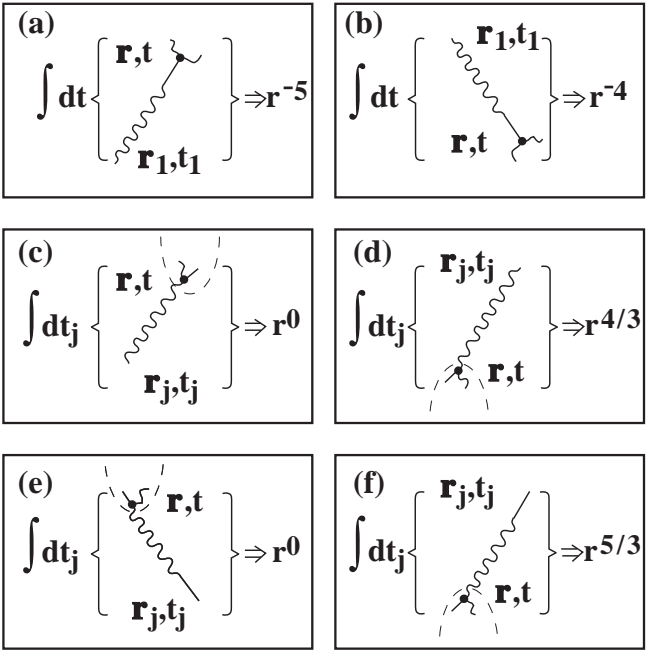}
 \vspace{.5cm}
 \caption{The elements that appear in the local geometries
 in Figs. 12 and 13.}
  \label{fig:fig14}
 \end{figure}

 The question that we want to answer is the following: when we integrate
 over $t, t { _{2}}$ and $t { _{3}}$ where is the dominant contribution to
 the integral over ${\bf r}$. In order to answer this question efficiently
 we decompose the various integrals to their elements, which are shown in
 Fig.\ref{fig:fig14}. Every vertex is a junction of two wavy lines and
 one straight line. The Green's function belonging to the latter is
 considered together with the vertex as one element, shown in panels a and
 b. The wavy line can belong to either a correlator considered in panels c
 and d, or to a Green's function considered in panels e and f. In panels
 c-f the vertex is excluded from considerations, as is indicated by the
 dashed line.  We evaluate the time integrals on the basis of (\ref{b13})
 and (\ref{b16}).  For brevity we will take $\alpha =\beta =0$.  This means
 that our results shown in Fig.13 are bounds, and the actual situation is
 always {\it better} in terms of the conclusion where the integral over
 ${\bf r}$ contributes most.

 The estimate of Fig.\ref{fig:fig14}a is $r { ^{-5}}$ which is obtained
 from the estimate of the vertex like $1/r$ and of the Green's function
 like $r { _{1}}/r { ^{4}}$.  The integral over t is restricted by $\tau(r
 { _{1}})$ and does not contribute.  Fig.\ref{fig:fig14}b is estimated as
 $r^{ -4}$ which stems from the estimate of  the Green's function as $1/r {
 ^{3}}$.  Fig.\ref{fig:fig14}c is $r { ^{0 }}$ because the correlator is $r
 { _{j}} { ^{2/3}}$ and the time restriction is $\tau(r { _{j}})$. On the
 contrary Fig.\ref{fig:fig14}d is $r { ^{4/3}}$ because both the correlator
 and the time restriction contribute $r { ^{2/3}} $ each.
 Fig.\ref{fig:fig14}e is again $r { ^{0 }}$ because the Green's function
 and the time restriction are determined by $r { _{j}}$.  On the contrary
 13f gives $r { ^{5/3 }}$ due to $r/r { _{j}} { ^{4 }}$ in the Green's
 functions and $r { ^{2/3 }}$ from the time restriction.

\begin{figure}
 \epsfxsize=8.6truecm
 \epsfbox{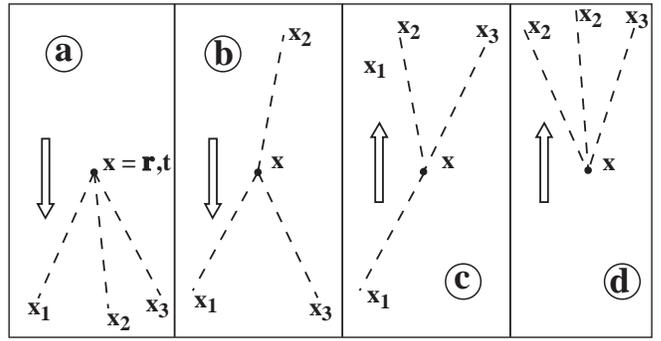}
 \vspace{.5cm}
 \caption{The final relevant geometries in stretched diagrams and the
         indication where the main contribution to the
         integral comes from.}
 \label{fig:fig15}
 \end{figure}

 To evaluate the integral over $x$ in Fig.\ref{fig:fig12}a we
 need to combine Fig.\ref{fig:fig14}a with
  Fig.\ref{fig:fig14}d and Fig.\ref{fig:fig14}f in all
  possible combinations. upon performing the integral over
  $r$ we see that we always have the contribution coming
  form the upper limit,i.e. when $r$ is of the order of $r {
 _{2}}$ or $r { _{3}}$. To evaluate Fig.\ref{fig:fig12}b we
 combine Fig.\ref{fig:fig14}a with Fig.\ref{fig:fig14}c or
 Fig.\ref{fig:fig14}e and with Fig.\ref{fig:fig14}d or
 Fig.\ref{fig:fig14}f. Here the resulting $r$ integral
 contributes mostly in the lower limit, when $r$ is or the
 order of $r { _{1}}$ or $r { _{3}}$.  Similarly we find
 that in 11c the major contribution is in the lower limit,
 as is the case for Fig.12c.  In Fig.\ref{fig:fig13}a and
 Fig.\ref{fig:fig13}b the major contribution is in the upper
 limit.  This information is summarized in
 Fig.\ref{fig:fig15} that shows where the integral over $r$
 contributes for any possible arrangement of propagators.
 The arrow shows whether the major contribution is coming
 from the upper or the lower limit. We again want to invoke
 the mechanical analogy of springs, and point out to the
 reader that one can guess where the major contribution
 comes from simply by counting how many springs pull in each
 direction. We note that the ``spring strength" is not the
 same for a correlator or a Green's function.  But in any
 combination two springs are always stronger than any single
 spring.

 These results can be used now to prove immediately that the simply
 decorated vertices in Fig.\ref{fig:fig6}  have their main contribution
 (in the stretched situation) in the region of integration $x { _{a}} \sim
 x { _{b}} \sim x { _{c}}$.  Consider for example the decorated vertex A in
 the diagram numbered as (1). Let us stretch the diagram such that $r {
 _{1}}$ and $r { _{2}}$ are the smallest positions and $r { _{3}}$ and $r {
 _{4}}$ the largest. Fix $r { _{b}}$ and realize that the results
 summarized in Fig.\ref{fig:fig15} imply that the main contribution
 is when $r { _{a}}$ is of the order of $r { _{b}}$. Similarly, the
 integral over $r { _{c}}$ will contribute in its lower limit, again for $r
 { _{c}}$ or the order of $r { _{b }}$.  Using now the proven locality $x {
 _{a}} \sim x { _{b}} \sim x { _{c}}$, the scaling relations (\ref{b11})
 immediately imply that the evaluation of the decorated vertex cannot be
 changed from the evaluation of the bare one, and it is therefore $1/r {
 _{b}}$.

  We could go on to any chosen decoration of the vertex and reach a similar
 conclusion. Consider for example the decoration of the same vertex A shown
 in Fig.\ref{fig:fig11}b . Fix the position $x { _{3}}$ for example.
 Stretch the springs connected to $x { _{a}}$ and $x { _{b}}$  such that $r
 { _{a }}<r { _{1 }}<r { _{3 }}<r { _{2}} <r { _{b}}$. The spring analogy
 which says that two springs are stronger than a single one implies that
 the dominant contribution comes from $r { _{1 }}$ of the order of the
 smaller of  $r { _{4}}$ and $r { _{6}}$. Similarly, $r { _{2}}$ will be of
 the order of the larger of $r { _{4}}$ and $r { _{5}}$. Next we repeat the
 argument at $r { _{4}}$ or $r { _{6}}$ and $r { _{4}}$  or $r { _{5}}$and
 conclude that they contribute in the vicinity of $r { _{3}}$. When all the
 coordinates are of the same order the scaling relations protect the
 evaluation of the dressed vertex and keeps it as $1/r { _{3}}$.  The
 spring analogy helps to understand that when we decorate the vertex
 further the rigidity is even more pronounced since there are more springs
 to keep it in place.

 \subsection{Rigidity of the Rungs}
 Having demonstrated the rigidity of the 3-point vertices, we can proceed
 now to examine the rigidity of the 4-point rungs.  For the case of the
 dressed simple ladders, the argument is immediate.  Consider the diagram
 in Fig.\ref{fig:fig8}a, and focus for example on the diagram designated
 as 1. Suppose that we stretched $r { _{1 }}$  and $r { _{2}}$ to be much
 smaller than $r { _{a}}$ and $r { _{b}}$, and $r { _{3}}$ and $r { _{4}}$
 to be much larger than $r_a$  and $r_b$. Can we have a significant
 contribution  from the region in which  $r { _{a}}$ becomes much smaller
 than $r { _{b}}$ or vice versa?  The analysis summarized in
 Fig.\ref{fig:fig15} or the spring analogy imply that the answer in
 negative.  Both vertices will contribute in the same neighborhood $r {
 _{a}}\sim r { _{b}}$. This is the property of rigidity of the dressed
 simple rung.

 Next we are going to consider the two-eddy mass operator $\Sigma_2 (
 {\bf r} { _{0}}|x { _{1}},x { _{2}},x { _{3}},x { _{4}})$ which is
 positioned in a ladder like in Fig.\ref{fig:fig9}b (diagram 3) with
 another such mass operator above and below. Here ``above" means that the
 lowest two coordinates of the next mass operator are much larger than $x {
 _{c}}$, $x { _{d}}$, and ``below" means that the upper two coordinates of
 the previous mass operator are much smaller than $x { _{a}}$, $x { _{b}}$.
 Consider this object with one fixed coordinate, say $r { _{a}}$, and
 integrate over the other three coordinates.  We will prove now that the
 main contribution to this integral comes from the regime in which all the
 three coordinates are of the order of the fixed one.

\begin{figure}
 \epsfxsize=8.6truecm
 \epsfbox{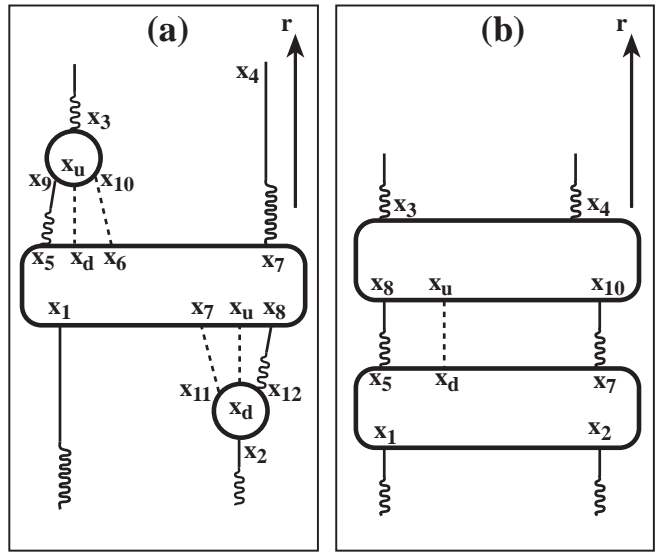}
 \vspace{.5cm}
 \caption{Diagrams used in the proof of rigidity of the rungs of the
         ladder.}
 \label{fig:fig16}
 \end{figure}

 To this aim consider Fig.\ref{fig:fig16}a. We display here a block
  representing a dressed rung of the ladder all of whose coordinates are of
 the same order of magnitude, except for the two blobs whose coordinates
 are much larger or much smaller than all the other coordinates.  These two
 blobs must be connected to the rung with at least two legs, else the
  diagram would be one-eddy reducible.

  Consider the case that the blob includes just one vertex, say $x {
 _{3}}$, and it is indeed connected via exactly two legs, say to the
 vertices $x { _{5}}$ and $x { _{6}}$. In this case we are in the situation
 of Fig.\ref{fig:fig15}b, and the vertex will contribute mostly when $x
 { _{3 }}$ is of the order  max$(x { _{5}}\,,\, x { _{6}})$. If the vertex
 gets decorated, the proof of rigidity of the 3-point objects guarantees
 that the dominant contribution to the integral comes from the regime in
 which the positions of all the vertices are of the same order, say $r {
 _{3}}\sim r { _{9}}\sim r { _{10}}$. The scaling relations imply that this
 vertex can be evaluated as, say, $1/r { _{3}}$, exactly like the bare one.
 Consequently the blob contributes mostly in the vicinity of max$(x {
 _{5}}\,,\,x { _{6}})$ again.

  Next we discuss the effect of adding additional legs to the connection of
 the blob to the main block of the rung. These legs can be either
 correlators $F( {\bf r} { _{0}}|x { _{u}},x { _{d}})$  or one of the
 Green's functions $G( {\bf r} { _{0}}|x { _{u}},x { _{d}})$ and $G( {\bf
 r} { _{0}}|x { _{d}},x { _{u}})$. Here $x { _{u}}$ belongs to the upper
 block $x { _{u}}\sim x { _{3}}$, and $x { _{d}}$ belongs to the lower
 block $x { _{d}}\sim x { _{5}}$. We will discuss the effect of the
 addition of one such leg, and show that it makes the integration over $x {
 _{3}}$ peak more strongly at its lower boundary.  The addition of more
 legs just enhances this tendency.

  Adding a correlator $F( {\bf r} { _{0}}|x { _{u}},x { _{d}})$ requires
 the addition of two vertices at $x { _{u  }}$ and $x { _{d}}$, two Green's
 functions, and two space time integrations over $x { _{u}}$ and $x {
 _{d}}$.  This leads to an additional dimensionless factor $(r { _{d}}/r
 { _{u}}) { ^{\gamma}}$. To evaluate $\gamma$ note that upstairs we have $r {
 _{u}} { ^{3 }}$ from the space integral, $1/r { _{u}} { ^{3}}$ due to the
 Green's function, and $1/r { _{u}}$ due to the vertex. The time integral
 upstairs contributes actually $\tau(r { _{d}})$ due to the time
 restriction on the correlator.  Downstairs we have $r { _{d}} { ^{3+2/3
 }}$ from the space time integral, $1/r { _{d}} { ^{4}}$ from Green's
 function and vertex, and $r { _{d}} { ^{2/3}}$ as an evaluation of the
 added correlator.  Consequently we find in this case $\gamma=1$.

  If the added leg is $G( {\bf r} { _{0}}|x { _{u}},x { _{d}})$ the
 situation upstairs changes since the time integral is now bounded by $\tau
 (r { _{u}},r { _{d}}) < \tau (r { _{u}})$. The situation downstairs also
 changes.  We need to insert an additional correlator (instead of Green's
 function) which contributes $r { _{d}} { ^{2/3}}$. The $G( {\bf r} {
 _{0}}|x { _{u}},x { _{d}})$ contributes $1/r { _{d}} { ^{3}}$, and the
 space time integral $r { _{d}} { ^{3+2/3}}$. By power counting we conclude
 that in this case $\gamma >1/3$.

 Finally the added leg can be $G( {\bf r} { _{0}}|x { _{d}},x { _{u}})$,
 which is evaluated as $r_d/r { _{u}} { ^{4}}$. In this case we add
 upstairs a correlator which contributes $r { _{u}} { ^{2/3}}$, and the
 space integral over the vertex gives $r { _{u}} ^3$. The time integral
 upstairs in bounded by $\tau (r { _{d}})$.  Downstairs we add a Green's
 function evaluated as $1/r { _{d}} { ^{3}}$, and the space time
 integration over the vertex contributes $r_{d}^{3+2/3}$. In
 summary we compute $\gamma > 4/3$.

 We see that in all cases $\gamma$ is positive, leading to a stronger
 preference of the integral for peaking downstairs, as claimed above. This
 result is again in conformity with the intuitive picture that considers
 the propagators as springs that pull fragments of a diagram to each other.
 Adding more propagators to the detached blob makes the situation even more
 rigid.

 The analysis of the blob with smaller coordinates parallels the above
 analysis exactly. Again the blob must be connected to the main rung with
 at least two propagators, and the spring analogy works perfectly.

 Finally consider the situation in Fig.\ref{fig:fig16}b, in
 which two blocks involving arbitrary resummations of
 two-eddy irreducible diagrams which are separated by at
 least three legs as shown. The two outer Green's function
 are parts of the principal paths, and the dashed line
 represents at least one additional connection, which is
 either $F( {\bf r} { _{0}}|x { _{u}},x { _{d}})$ or one of
 the Green's functions $G( {\bf r} { _{0}}|x { _{u}},x {
 _{d}})$ and $G( {\bf r} { _{0}}|x { _{d}},x { _{u}})$. If
 all the coordinates were of the same order, this diagram
 would represent one rung of the ladder, and its evaluation
 would be, according to the scaling relations, just like the
 evaluation of a bare rung. The aim of the present
 discussion is to show that indeed in the stretched diagram the largest
 contribution comes from the regime in which all the coordinates are of the
 same order.  To show this we will assume that the two blocks have widely
 separated coordinates. All the upper coordinates are of the order of $x {
 _{u}}$, and all the lower coordinates are of the order of $x { _{d}}$. We
 can fix one coordinate at will, say $x { _{1}}$. The claim is that the
 major contribution to the diagram in Fig.\ref{fig:fig16}b comes from the
 regime in which all the coordinates are of the order of $x { _{1}}$.

  For pedagogical purposes it is useful to discuss first the two-eddy
 reducible topology in which there is no connection (in addition to the two
 Green's functions belonging to the two principal paths) between the two
 sub-blocks. If we assert that the coordinates in each block separately are
 of the same order, $x { _{u}}$ and $x { _{d}}$ respectively, the
 integration over all coordinates will result in a term proportional to
 $\int _{r_{d}} dr { _{u}}/r { _{u}}$.  Analytically this estimate is
 understood from the discussion of the bare ladder in Sect.\ref{19-sect3}b
 and from the rigidity of the dressed rung. In the spring analogy we have
 two springs below and two springs above, and this results in a neutral
 situation which is a ``mechanical balance".  Analytically this means that
 the integral is logarithmic, having no preference to either upper or lower
 bound. Adding any one leg that turns these two sub-blocks into a two-eddy
 irreducible topology immediately introduces into this integral a factor
 $(r { _{d}}/r { _{u}}) { ^{\gamma}}$ a with $\gamma$ taking one of the
 positive values found above.  This factor forces the integral to peak in
 the regime $r { _{u}}\sim r { _{d}}$. Any additional leg connecting the
 two sub block only serves to enhance this tendency. The additional spring
 turns the whole rung rigid.

 One can come up with arbitrarily involved topologies of sub-blocks, say
 by introducing additional blobs at various coordinates between the two
 blobs shown in Fig~\ref{fig:fig16}b. The analysis presented here suffices
 to show that all these cases lead to the conclusion that the major
 contribution comes from the shell in space in which all the coordinates
 are of the same order.

 \subsection{Dangerous contributions: ladders in ladders.}
 The procedure outlined above consists of a proof of rigidity in every
 order of the diagrammatic representation for the two-eddy irreducible mass
 operator which appears as a rung in the ladder.  However, the anomalous
 exponents discussed above appears in a power law that results from the
 resummation of ladder diagrams to all orders. Therefore we need to be
 doubly careful about the appearance of resummations to all order {\it
 within} the blocks that represent the rungs in our ladder.  In other
 words, we need to examine the possible appearance of ladders within the
 rungs of the ladders.

\begin{figure}
 \epsfxsize=8.6truecm
 \epsfbox{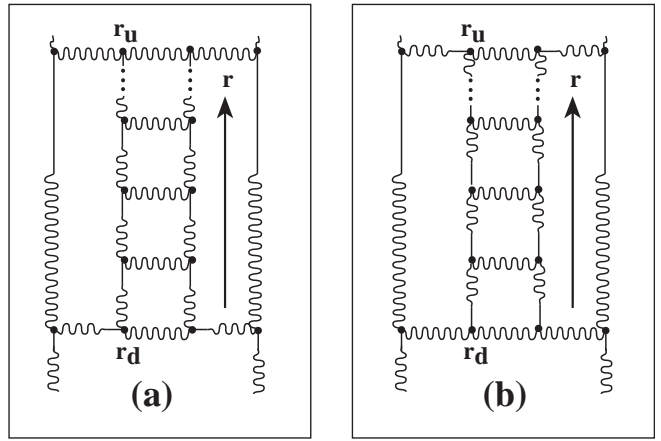}
 \vspace{.5cm}
 \caption{Examples of dangerous ``ladders within the ladder" diagrams.}
 \label{fig:fig17}
 \end{figure}

\begin{figure}
 \epsfxsize=8.6truecm
 \epsfbox{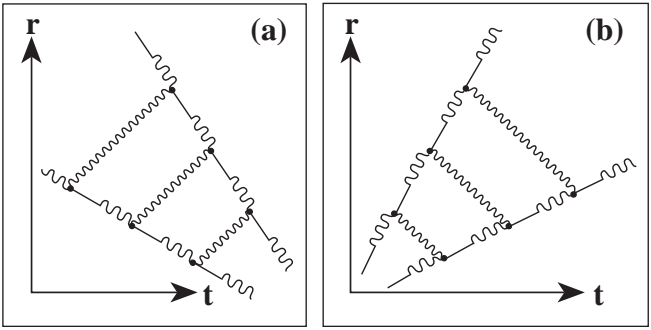}
 \vspace{.5cm}
 \caption{Ladders drawn in {\bf r},t coordinates, to show the natural
         leaning over that is dictated by causality.}
 \label{fig:fig18}
 \end{figure}

 In Fig.\ref{fig:fig17} we present examples of two dangerous ladders in
 the rung of the ladder. In panel (a) the inner ladder has Green's
 functions  oriented in the same sense as the main ladder. In panel (b) the
 Green's function are oriented in the opposite sense.  In
 Figs.\ref{fig:fig18}17a and \ref{fig:fig18}b we present the inner ladders
 in $r,t$ coordinates, to stress that the time restrictions on the Green's
 functions force the two inner ladders to be oriented in the opposite
 direction in time.

  The understanding of the situation of Fig.\ref{fig:fig17}a is not
  difficult.  We have learned before that the insertion of a leg of $G(
 {\bf r} { _{0}}|x { _{d}},x { _{u}})$ results in a reduction factor of the
 order of $(r { _{d}}/r { _{u}}) { ^{4/3}}$. The inner ladder in Fig.16a
 starts with two such legs, and this results in a total reduction of the
 order of $(r { _{d}}/r { _{u}}) { ^{8/3}}$. On the other hand, the ladder
 itself will contribute an additional factor of $(r_{u}/r_{d})^{\Delta}$.
 We thus expect that in toto we will have a factor of $(r { _{d}}/r {
 _{u}}) ^{8/3-\Delta}$.

 This calculation show that if $\Delta$ exceeds $8/3$ we will lose the
 property of rigidity and our theory will be in real danger. In fact we
 will show momentarily that the borderline of applicability of the theory
 is actually $\Delta=4/3$. As long as we assume that the numerical value of
 $\Delta$ is smaller than  4/3 we can conclude that the inner ladder still
 acts as a spring which is at least as strong as an additional $G( {\bf r}
 { _{0}}|x { _{d}},x { _{u}})$ leg. It is in no way dangerous.

 The understanding of the role of the second type of ladder,
 Fig.\ref{fig:fig17}b and \ref{fig:fig18}b,  calls for a different type of
 consideration. Suppose that we want to insert the ladder of 17b between
 two rungs (say, between the lowest and the next lowest rungs) of an outer
 ladder of the type 17a. However 17b is oriented (by causality) upward to
 the right. Since the ladder in the 17a is oriented (again by causality)
 upward to the left, inserting 17b between two rungs severely constrains
 the time range of the inserted ladder. This time constraint prevents a
 logarithmic divergence. In other words we cannot have a large factor of
 $(r { _{u}}/r { _{d)}}$ to some exponent in the inserted ladder.  The
 lesson is that consequences of causality on the time restriction of the
 Green's functions do not allow an insertion of one type of ladder in the
 other.

\begin{figure}
 \epsfxsize=4.5truecm
 \epsfbox{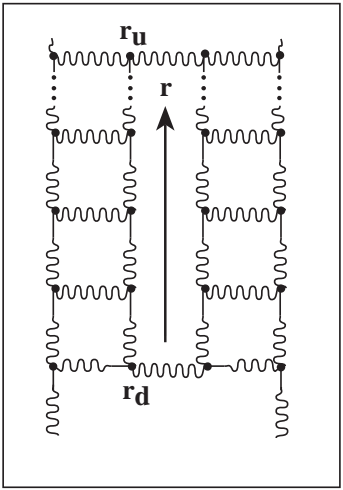}
 \vspace{.5cm}
 \caption{More dangerous ``ladders within the ladder" diagram.}
 \label{fig:fig19}
 \end{figure}

 However these are not the only dangerous situations that need
 further careful discussion. For example in Fig.\ref{fig:fig19} we
 display yet another configuration of contributing diagrams in which the
 bare Green's functions that define the ladder are replaced themselves by
 ladders. In this case the estimate of the case shown in Fig.~16a is
 corrected by another anomalous factor of $(r { _{u}}/r { _{d}})
 ^{\Delta}$.  One can still state that for $\Delta < 4/3$ the power $-8/3$
 keeps the rung rigid, but we already come closer to a possible breakdown
 of the property of rigidity which would occur if  $\Delta = 4/3$. Indeed,
 we are going to show below that $\Delta$ attains precisely its critical
 value $2-\zeta_2$ which for K41 scaling is exactly 4/3. This fact means
 that nonperturbative effects which appear in infinite resummations are
 very important, and they may indeed lead to nontrivial renormalizations
 of the scaling exponents.
 \subsection{Radius of the ball of locality}
 \label{19-sect5}
 Before continuing our study of anomalous scaling we pause to make use of
 the concept of rigidity to improve our estimate of the radius of the ball
 locality. The question arises in the context of the Green's function
 $G(0|{\bf r},{\bf r}',t)$ and the correlation function
 $F(0|{\bf r},{\bf r}',t)$ when $r$ and $r'$ are of different order of
 magnitude. The result of this consideration will be that the radius of
 the ball of locality is always determined by the smaller of $r$ and
 $r'$. The reader who is not interested in this issue is invited to go
 directly to Sect.\ref{19-sect6}.

 In the context of the Green's function the question is what is the
 spatial domain which contributes mostly to the integration over $\bf r_1$
 and $\bf r_2$ in Eq.(\ref{b8}). Due to rigidity, in the stretched
 situation when $r\ll r'$ or vice versa the coordinates $r_1$ and $r_2$
 must be of the same order to contribute significantly. In this regime the
 mass operator $\Sigma$ can be evaluated as (cf. paper I) $S_2(r_1)/r_1^5$
 $\sim \bar\varepsilon ^{2/3} r_1^{-13/3}$. The integrals over ${\bf
 r_1}$, ${\bf r_2}$  and $t_1$  are bounded $\int r_1^5 \tau (r_1) dr_1$.
 Next the product of the two remaining Green's function can be always
 estimated as  $G(0|{\bf r},{\bf r}',0^+)/r_1^3$.  The total $r_1$
 dependence of the integrand is $r_1^{-4/3}$ and therefore the integral
 contributes at its lower limit which is the smaller of $r$ and $r'$.

 In the context of the correlator we need to analyze the integrals over
 ${\bf r_1}$, ${\bf r_2}$, $t_1$  and $t_2$ in Eq. (\ref{b9}). In the
 stretched situation rigidity allows us to restrict our considerations to
 the regime $r_1 \sim r_2$.  In this regime the mass operator $\Phi$ can be
 evaluated (cf. I) as $S_2^2(r_1)/r_1^2$.  The space-time integrals
 contribute to the $r_1$ dependence as before, because one of the time
 restrictions is $\tau(\min \{r,r'\})$. Now the two Green's functions are
 differently oriented and can be evaluated as $\min \{r,r'\}/r_1^7$. In
 toto we find the same evaluation of the integrand in (\ref{b9}),
 namely $r_1^{-4/3}$. Again the integral contributes at its lower limit
 which is the smaller of $r$ and $r'$.

 The conclusion is that due to rigidity, in both Eqs.  (\ref{b8},\ref{b9})
 the relevant domain of integrations over ${\bf r_1}$ and ${\bf r_2}$  is
 $r_1 \sim r_2 \sim \min \{r,r'\}$.  Now we can apply the property of
 locality proven in I to see  that upon expanding the mass operators in
 their infinite series, {\it all} the intermediate coordinates have to
 belong to the same domain. Consequently, the global ball of locality has a
 radius $\min \{r,r'\}$.
 \section{Anomalous scaling}
 \label{19-sect6}
 In this section we discuss the resummed equation for the nonlinear Green's
 function, and solve for the anomalous exponent that is
 associated with this function.  The equation is shown graphically in Fig.
 \ref{fig:fig10}b, and in analytic form it reads
 \begin{eqnarray}
 &&
 \!\!\!\!\!\!
 G  _{2}^{\alpha\beta\gamma\delta}
 ( {\bf r} { _{0}}|x { _{1}},x { _{2}},x { _{3}},x { _{4}}) \equiv
 G ^{\alpha\gamma}( {\bf r} { _{0}}|x { _{1}},x { _{3}})
 G ^{\beta\delta}
 ( {\bf r} { _{0}}|x { _{2}},x { _{4}})
 \nonumber \\
 &+& \int dx { _{a}}dx { _{b}}dx { _{c}}dx { _{d}}
 G ^{\alpha\mu}( {\bf r} { _{0}}|x { _{1}},x { _{a}})
 G ^{\beta\nu}( {\bf r} { _{0}}|x { _{2}},x { _{b}})
 \label{e1}
  \\
 &\times& \Sigma_{2}^{\mu\nu\sigma\rho}
 ( {\bf r} { _{0}}|x { _{a}},x { _{b}},x { _{c}},x { _{d}})
 G_{2\,\sigma\rho\gamma\delta}
 ( {\bf r} { _{0}}|x { _{c}},x { _{d}},x { _{3}},x { _{4}})
 \nonumber
 \end{eqnarray}
 Eq. (\ref{e1}) is a closed integral equation for ${\bf G}_2$, but the
 operator $\Sigma { _{2}}$ is given in terms of an infinite series which
 begins with the diagrams shown in Fig.\ref{fig:fig10}a.

 Equation (\ref{e1}) can be considered as a linear inhomogeneous equation for
 $G  _{2}$. If we expand around the inhomogeneous term $G\times G$ we recover
 the initial expansion that was represented diagrammatically above. However,
 we can now seek nonperturbative solutions which are the solutions of the
 homogeneous part of the equation (\ref{e1}). We will demonstrate
 in subsection A that
 this nonperturbative solution has a power law form in which the anomalous
 exponent $\Delta$ appears, and that this solution is much larger than the
 inhomogeneous solution. In subsection B we will evaluate the anomalous
 exponent exactly. In order to define the anomalous exponent $\Delta$
 precisely we introduce the following function:
 \begin{equation}
 T_{\alpha\beta}({\bf r}_1,{\bf r}_2, R) \equiv \int dx_3dx_4
 G  _{2}^{\alpha\beta\gamma\delta}
 ( {\bf r} { _{0}}|x { _{1}},x { _{2}},x { _{3}},x { _{4}}) D^{^{(R)}}_
 {\gamma\delta}(x_3,x_4) \ , \label{defT}
 \end{equation}
 where $ D^{^{(R)}}_{\gamma\delta}(x_3,x_4)$ is  some function with a
 characteristic length scale $R$ and characteristic time scale $\tau(R)$.
 The anomalous exponent is defined via the limit
 \begin{equation}
 \lim_{\eta<r_1,r_2\ll R}\bbox{\nabla}_1\cdot\bbox{\nabla}_2 T_{\alpha\alpha}
 ({\bf r}_1,{\bf r}_2,R) \propto {1\over r_{12}^{\Delta}} \ .\label{defdel}
 \end{equation}
 The reasons for this somewhat cumbersome definition of $\Delta$ will become
 clearer in paper III of this series.
 \subsection{Homogeneous Solutions: Qualitative Analysis}
  The equation (\ref{e1}) is not an easy equation to solve. The nonlinear
 Green's function is a function of four space-time variables, and it is a
 fourth-rank tensor. In order to gain insight on ${\bf G}_2$ we will reduce
 it to a function of a smaller number of variables. Also, we are interested
 here in its scaling properties only, and we can simplify the discussion by
 dropping the tensor indices. This will allow us to develop a qualitative
 analysis that will demonstrate the power law behaviour that is implied in
 (\ref{defdel}). The actual computation of the limit in (\ref{defdel}) will
 be done in the next subsection.

  The reduction in number of variables is done as follows: we integrate
 ${\bf G}_2$ over the last two time variables $t { _{3}}$ and $t { _{4}}$,
 and consider its value at $t { _{1}}=t { _{2}}=0$ and ${\bf r} { _{1}}=
 {\bf r} { _{2}}= {\bf r}$, and ${\bf r} { _{3}}= {\bf r} { _{4}}= {\bf
 R}$.  We choose $R \gg r$.  Define a new quantity $\tilde g_{2}
 ( {\bf r}, {\bf R})$ via the equation
 \begin{eqnarray}
 &&\int dt { _{3}}dt { _{4}} G_2( {\bf r} { _{0}}| {\bf r},t { _{1}}=0,
 {\bf r},t { _{2}}=0, {\bf R},t { _{3}}, {\bf R},t { _{4}}))
 \nonumber \\
 &\equiv& \big[\tau(r,R) G { ^{0}}( {\bf r} { _{0}}| {\bf r}, {\bf R},
 0 { ^{+}})\big] { ^{2}} \tilde g _{2}
 ( {\bf r}, {\bf R})\  .
 \label{e4}
 \end{eqnarray}
 The function $\tilde g _{2}( {\bf r}, {\bf R})$ is a dimensionless
 function by construction. In general $\tilde g _{2}( {\bf r}, {\bf R})$
 is a function of $r$, $R$ and the angle between ${\bf r}$ and ${\bf R}$.
 For $R\gg r$ the angle becomes irrelevant. In the regime when $r$ and $R$
 are in the inertial interval, the function  $\tilde g _{2}$
 can depend on the ratio $r/R$ only. The notation will be
 \begin{equation}
 y=r/R \,, \quad \tilde g _{2}(r,R) = g { _{2}}(y) \ . \label{g_2}
 \end{equation}
 In Appendix A we show that this function
 satisfies the approximate integral equation
 \begin{equation}
 g { _{2}}(y)  = C\int { dy { _{a}} \over
 y { _{a}}} K(y,y { _{a}}) g { _{2}}(y { _{a}})
 \label{e12}
 \end{equation}
 with  some dimensionless constant $C$, and kernel $ K(y,y { _{a}})$
 which can be written as
 \begin{equation}
 K(y,y _a) =\left[{ g(y/y { _{a}}) g(y { _{a}})
 \over g(y) } \right]^{2}\ .
 \label{e13}
 \end{equation}
 Here $g(y)$ and $g_2(y)$ are  dimensionless functions which are defined
 by
 \begin{eqnarray}
 g(r/r') &\equiv&  r' { ^{3}}G( {\bf r} { _{0}}|r,r',0 { ^{+}})\,,
 \label{e14}   \\
 g_2(r/r') &\equiv& \tilde g_2(r,r') \ .
 \nonumber
 \end{eqnarray}

 Eq.(\ref{e12}) was derived on the basis of the choice of $y\ll 1$ and
 considering $y \ll  y { _{a }}\ll  1$. Of course, the equation for the
 nonlinear Green's function included additional regimes that we did not
 study explicitly. To gain insight to the relevance of these regimes we are
 going to interpret Eq.(\ref{e12}) in a more general setting, allowing $y$
 and $ y { _{a}}$ to go between zero and infinity:
 \begin{equation}
 g { _{2}}(y)  = A\int _0^\infty { dy_{a} \over y_{a} }
  K(y,y { _{a}}) g { _{2}}(y { _{a}}) \ .
 \label{e15}
 \end{equation}
 In this subsection we examine the solutions of the model Eq.(\ref{e15}).

 We note that the known asymptotic properties of $G( {\bf r} { _{0}}|r,r',0
 { ^{+}})$ imply that
 \begin{equation}
 g(y) =  \cases   { a\quad &{\rm for}\quad $ y\gg 1$\,,    \cr
                    b\,y \quad &{\rm for}\quad $ y\ll 1$ \ . }
 \label{e16}
 \end{equation}
 with $a$ and $b$ being dimensionless coefficients. These properties
 imply also the asymptotic properties of the kernel $K(y,y { _{a}})$. In
 the $y$, $y { _{a}}$ plane we display the evaluation of the kernel in Fig.
 \ref{fig:fig20}.

 In order to understand the type of solutions that are supported by the
 integral equation (\ref{e15}) we turn now to some simplified models.

 \subsubsection{Model A}
 The simplest model that we can think of is the one in which the constant
 $a$ in Eq.(\ref{e16}) is zero. This leaves us with $g(y)$ in the regime
 $y<1$ where our model was derived.  The asymptotic form of $K(y,y {
 _{a}})$ is therefore $K(y,y { _{a}}) = 0$ for $y { _{a }}>1$ and for $y {
 _{a }}< y$.  In other words, $K(y,y { _{a}})=b { ^{2}}$ in region 2 of
 Fig.\ref{fig:fig20}, and zero in all the other regions. We call this model
 A.  In this case the integration in (\ref{e15}) is the trajectory denoted
 by the dashed line A$_{1}$ in Fig.\ref{fig:fig20}, which is limited to lie
 within region 2. The equation for $g { _{2}}(y)$ reads
 \begin{equation}
 g { _{2}}(y) = {\Delta}{ _{0}}\int_y^1{ dy { _{a}}\over y_{a} }
 g { _{2}}(y { _{a}})\ .
 \label{e17}
 \end{equation}
 where $\Delta _{0}= A b{ ^{2}}$. The solution of this equation is
 \begin{equation}
  g { _{2}}(y) = C/ y ^{\Delta} \quad
 \Delta =\Delta { _{0 }}\quad {\rm (model\ A)}\ .
 \label{e18}
 \end{equation}
 The coefficient $C$ should be determined from the boundary condition
 $ g { _{2}}(1)$. In dimensional form this result reads
 \begin{eqnarray}
 && \int dt _{3} dt  _{4} G( {\bf r} _{0} | {\bf r},t { _{1}}=0,
 {\bf r},t { _{2}}=0, {\bf R},t { _{3}}, {\bf R},t { _{4}}))
 \nonumber \\
 &=&C\,
 [\tau(r,R) G^{0} ( {\bf r} _{0} | {\bf r}, {\bf R},0^{+})]^2
 \left({R\over r}\right )^{\Delta_0}\ .
 \label{e19}
 \end{eqnarray}
  This is the same result that we obtained in section 3.B on the basis of
 the resummation of the logarithmic contributions in the simple ladder
 diagrams. This is not surprising since model A represents exactly in
 resummed form the nature of the approximation in section 3.B in which
 only the dominant contribution was taken from each diagram. The
 subdominant contributions are not expected to ruin the anomalous scaling
 behavior, but since the anomalous exponent is sensitive to the numerical
 value of the coefficients, we expect the subdominant terms to effect the
 numerical value of the exponent. To study this effect we turn to a
 slightly more complicated model.
 \subsubsection{Model B}
\begin{figure}
 \epsfxsize=7truecm
 \epsfbox{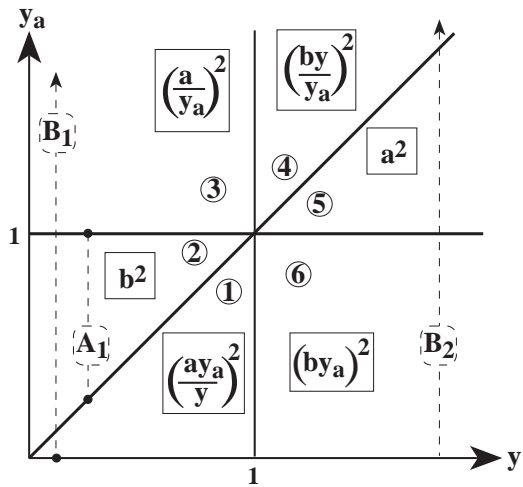}
 \vspace{.5cm}
 \caption{The kernel of models A (trajectory A$_1$) and B (trajectories
 B$_1$ and B$_2$).}
 \label{fig:fig20}
 \end{figure}

 The next model, which we refer to as model B, is obtained when we use the
 asymptotic properties of $K(y,y { _{a}})$ as shown in Fig.\ref{fig:fig20}
 up to the boundaries of the regions, choosing the coefficients $a=b$.
 Start with $y < 1$, and follow the trajectory B$_{1}$ of
 Fig.\ref{fig:fig20} in the integration in Eq.  (\ref{e15}).  This
 trajectory crosses three regions in Fig.\ref{fig:fig20}, namely 1,2 and 3,
 and accordingly the equation has three integrals:
 \begin{eqnarray}
 g{ _{2}}(y) &=& {\Delta}_0 \Bigg[{1\over y^2}
 \int _0^y dy { _{a}} y { _{a }}g { _{2}}(y { _{a}})
 \label{e20}  \\
 &+& \int_y^1{ dy { _{a}}\over y_{a} }
  g { _{2}}(y { _{a}})) +\int _1^\infty {dy { _{a }}\over y_{a}^3 }
  g { _{2}}(y_a) \Bigg]\,, \  y<1
 \nonumber
 \end{eqnarray}
 The second integral is the one taken into account in Model A.  We seek a
 solution in the form of (\ref{e18}). Substituting this form in
 (\ref{e20}) we find that the requirement of convergence of the integrals
 puts a limit on the allowed values of $\Delta$. These limits are
 \begin{equation}
 -2 <{\Delta}< 2 \ .
 \label{e21}
 \end{equation}
 The  two solutions  for ${\Delta}$ are
 \begin{equation}
 \Delta_{\pm} = 1\pm \sqrt{1-2\Delta_0} \quad{\rm  (model\ B}\,,\
 y<1{\rm )}  \ .
 \label{e22}
 \end{equation}

\begin{figure}
 \epsfxsize=7truecm
 \epsfbox{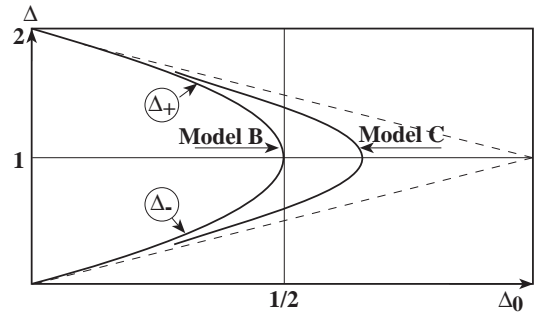}
 \vspace{.5cm}
 \caption{The solution {$\delta$} as a function of {$\Delta_0$}
          in model B and model C.}
  \label{fig:fig21}
 \end{figure}

 The two real branches of solutions as a function of $\Delta_{0}$ are shown
 in Fig.\ref{fig:fig21}. For ${\Delta}{ _{0}} \ll 1$  the solution
 ${\Delta}{ _{- }}$ of model B coincides with the solution of model A.
 Indeed, for small ${\Delta}{ _{0}}$ the second integral in (\ref{e19}) is
 dominant since it is the only one proportional to $1/\Delta$. This
 contribution is equivalent to considering the ladders in their fully
 stretched configuration, which corresponds to model A. However in this
 case there is another branch of solutions, ${\Delta}{ _{+}}$, which for
 small ${\Delta}{ _{0}}$ is dominated by the first integral in (\ref{e20})
 which is proportional to $1/(2-\Delta)$.  For ${\Delta}{ _{0}} = 1/2$
 these two branches coincide.  In this model there are no real solution for
 ${\Delta}{ _{0}} > 1/2$.

  Since we have two solutions for $\Delta$, the  general solution is a sum
 \begin{equation}
 g { _{2}}(y) ={C_{+} \over y ^{\Delta_+}}
 +  { C { _{-}}\over y ^{\Delta_-}}\, ,
 \label{e23}
 \end{equation}
 with coefficients $C { _{+}}$ and $C { _{-}}$ that are determined by the
 boundary condition and the requirement of continuity across the boundaries
 of Fig.26.

  We see that in the regime $y<1$ the solution (\ref{e23}) has a
 leading scaling exponent which is $\Delta{ _{+}}$, whereas $\Delta{ _{\_}}$
 appears only as a correction to scaling . In our simple model the leading
 scaling exponent $\Delta{ _{+ }}$ takes on values between 2 and 1 before
 becoming complex. In general we may have a whole spectrum of exponents
 that appear as corrections to scaling.

 \subsubsection{ Model C}
 Finally, we discuss for completeness a model C that introduces yet more
 freedom in the scaling function $g(y)$ :
 \begin{equation}
 \frac{g(y)}{a} = \cases {1\,\quad &  {\rm for}\ \ \ $  y>1/d $\,,  \cr
                     \sqrt{d\,y} \quad & {\rm for}\ \ \ $ d<y<1/d\,, $  \cr
                     y  \quad & {\rm for}\ \ \ $  y<d\ . $ }
 \label{e27}\\
 \end{equation}
 with $d < 1$ (for $d=1$ we regain model B). Model C introduces a smoother
 cross over in the regime $y \sim 1$. Substituting (\ref{e27}) in the
 expression for $K(y,y { _{a}})$ (\ref{e13}) we find for $y \ll  d$  the
 following expressions
 \begin{equation}
 {K(y,y_a)\over a^2}= \cases{
  \big( y_a/y \big) ^{2} &  {\rm [reg.}\ 1:\ \ \ $y { _{a }}< dy $\ ]\,,\cr
   d y_{a}/ y  & {\rm         [reg.}\ 1-2:\ $ dy < y { _{a}} <y/d$\ ]\,,\cr
   1 & {\rm [reg.}\ 2:\ \ \ $  y/d < y{ _{a}} <d$\ ]\,, \cr
   d/ y_a  & {\rm [reg.}\ 2-3: \ \ $d<y { _{a}} <1/d$]\,, \cr
  1/y_a^2  & {\rm [reg.}\ 3:\ \ \ $ 1/d < y$]\ .}
 \label{e28}
 \end{equation}
 The expression for $K(y,y { _{a}})$ in the regions 1,2,3 corresponds to
 the values in model B, cf. Fig.\ref{fig:fig20}.  For $d=1$ the regions 1-2
 and 2-3 disappear.  Now the integral equation for $g_2(y)$ takes the form
 \begin{eqnarray}
 g { _{2}}(y) &=& \Delta{ _{0}}\Bigg[ {1\over y^2}\int_0^{d\,y} dy {
 _{a}} y { _{a }}g { _{2}}(y { _{a}})
 \nonumber \\
 &+&{d\over y}\int_{d\,
 y}^{y/d} dy { _{a }}g { _{2}}(y { _{a}}) +\int_{y/d}^d {dy { _{a}}\over y
 { _{a}}} g { _{2}}(y { _{a}})
 \label{e29} \\
 &+&d\int _d^{1/d}{dy { _{a}}\over y { _{a}} { ^{2}}} g { _{2}}(y { _{a}}))
 +\int_{1/d}^{\infty}{dy { _{a }}\over y { _{a}} { ^{3}}} g { _{2}}(y {
 _{a}})\Bigg]\   .
 \nonumber
 \end{eqnarray}
 The requirement of convergence is again $-2<{\Delta}< 2$, since it is
 coming from regions 1 and 3 which are the same as in model B.  Seeking a
 solution in the form (\ref{e18}) we find that ${\Delta}$ solves the
 transcendental equation
 \begin{equation}
 { d^{2-\Delta}\over    2-\Delta}
 + {d^{\Delta}\over \Delta} +{ d^{\Delta}-d^{2-\Delta} \over 1-\Delta}
   = {1\over \Delta_{0} }\  .
 \label{e30}
 \end{equation}
 One obvious property of the solution of this equation is that the
 symmetry ${\Delta}\to (2-\Delta)$ is preserved, meaning that the solution
 is still symmetric around the ${\Delta}= 1$ line. Secondly, in the regime
 $\Delta { _{0 }}\to 0$ we have again the same two solutions as in
 model B i.e. ${\Delta}= \Delta { _{0}}$ and ${\Delta}=2 -\Delta { _{0}}$.
 The solution crosses the symmetry line $\Delta =1$ just once, exactly when
 \begin{equation}
 \Delta { _{0}} = 1/ 2(d+\ln\, d)\  .
 \label{e31}
 \end{equation}
 The implication of this result is that the topology of the line of
 solutions remains similar to the solution of model B, as shown in
 Fig.\ref{fig:fig21}.
 In particular the leading anomalous exponent is again larger than 1.


 \subsection{Exact Solution of the Anomalous Exponent $\Delta$}

 In this subsection we derive the  important exact result
 \begin{equation}
 \Delta = 2 -\zeta_2 \ . \label{delta_c}
 \end{equation}
 In order to prove this we will first establish an identity
 which relates the mass operator $\Phi$ which is shown in Fig.2b and the
 two-eddy irreducible mass operator $\Sigma_2$ which is expanded in Fig.10a.
 The identity is
 \begin{equation}
 \Sigma_2^{\alpha\beta\gamma\delta}({\bf r}_0|x_1,x_2,x_3,x_4) =
 {\delta\Phi_{\alpha\beta}(x_1,x_2)\over \delta F^{\rm (pc)}_{\gamma\delta}
 (x_3,x_4)} \ , \label{ident}
 \end{equation}
 where the superscript ``(pc)" means that the functional derivative is
 taken only with respect of the correlators on the principal cross section
 of the diagrams of $\Phi$.

 The proof of the identity is available from inspection of Figures 2 and
 10. By taking the functional derivative of diagram (1) in Fig.2 we get the
 first contribution to the first diagram on the RHS of Fig.10a (The one
 with two A vertices).  The factor of 1/2 in Fig 2 disappears because there
 are two identical contributions to the functional derivative. If we take
 all the diagrams in $\Phi$ with just two correlators in the principal
 cross section, then the functional derivative will yield exactly the first
 term on the RHS of Fig.10a. Next consider diagram (2) with three wavy
 lines at the cross section. This is the first in an infinite series of
 diagrams with three wavy lines at the cross section. Finding the
 functional derivatives of these contributions we generate all the terms in
 $\Sigma_2$ with two wavy lines at the cross section. For example, the
 diagram (2) in Fig.2 produces three terms, the second, third and fourth in
 Fig.10 with bare vertices. The generalization of the  procedure is clear.

 Next we consider the Wyld equation (\ref{b9}), and evaluate the functional
 derivative $\delta F_{\alpha\beta}({\bf r}_0|x_1,x_2)/ \delta D^{\rm
 (pc)}_{\gamma\delta}(x_3,x_4)$ where the functional derivative has the
 same meaning as Eq.(\ref{b6}), but we restrict the variation only to
 contributions appearing in the principal cross section in the diagrammatic
 representation of $F$.  Rewrite the Wyld equation in schematic form
 \begin{equation}
 F=G*[D+\Phi]*G \label{wyldsimp}
 \end{equation}
 where the star operation $*$ means integration over space and summation
 over the tensor indices. In the same schematic representation the
 functional derivative takes the form
 \begin{equation}
 {\delta F\over \delta D^{(pc)}} = GG +G*{\delta \Phi\over \delta D^{(pc)}}
 *G \ .
 \label{funder}
 \end{equation}
 Next observe that
 \begin{equation}
 {\delta \Phi\over \delta D^{\rm (pc)}}={\delta \Phi\over \delta F^{(pc)}}*
 {\delta F\over \delta D^{\rm (pc)}} \ . \label{chainrule}
 \end{equation}
 Using now the identity (\ref{ident}) we conclude that $\delta F/ \delta
 D^{\rm (pc)}$ solves exactly the same integral equation (\ref{e1}) as the
 nonlinear Green's function, i.e the equation displayed in Fig.10b. This
 means that
 \begin{equation}
 {\delta F_{\alpha\beta}({\bf r}_0|x_1,x_2)\over \delta D^{\rm (pc)}_
 {\gamma\delta}(x_3,x_4)} = G_2^{\alpha\beta\gamma\delta}({\bf r}_0|x_1,x_2,
 x_3,x_4) \ . \label{relation}
 \end{equation}
 One should note that if we were evaluating the full functional derivative
 of $F$ with respect of $D$, not restricted to the principal cross section,
 we would have derived another relation, which is
 \begin{equation}
 {\delta F_{\alpha\beta}({\bf r}_0|x_1,x_2)\over \delta D_
 {\gamma\delta}(x_3,x_4)} = \left<{\delta^2 w_\alpha({\bf r}_0|x_1)
   w_\beta({\bf r}_0|x_2)\over \delta h_{\gamma}(x_3)\delta h_{\delta}(x_4)}
 \right> \ . \label{newG}
 \end{equation}
 Here the RHS is another type of nonlinear Green's function.

 At this point we can use the result (\ref{relation}) in Eq.(\ref{defT})
 to obtain the relation
 \begin{equation}
 T_{\alpha\beta}({\bf r}_1,{\bf r}_2, R) \equiv \int dx_3dx_4
 {\delta F_{\alpha\beta}({\bf r}_0|x_1,x_2)\over \delta D^{\rm (pc)}_
 {\gamma\delta}(x_3,x_4)} D^{^{(R)}}_
 {\gamma\delta}(x_3,x_4) \ . \label{Tcalc}
 \end{equation}
 Now the physical meaning of the function $T_{\alpha\beta}({\bf r}_1, {\bf
 r}_2,{\bf R})$ becomes clear. It is the the change in the double
 correlator $F_{\alpha\beta}({\bf r}_1,{\bf r}_2)$ due to the existence of
 an additional random forcing on the scale $R$ with a correlation
 $D^{^{(R)}}$. The point now is that if the coordinates ${\bf r}_1$ and
 ${\bf r}_2$ are much smaller than $R$, then the double correlator should
 have its usual universal exponent $\zeta_2$  and the role of the
 additional forcing is only in changing the magnitude of $F$. Accordingly
 \begin{equation}
 \lim_{r_1,r_2\ll R}T_{\alpha\beta}({\bf r}_1,{\bf r}_2, R)\propto F_
 {\alpha\beta}({\bf r}_1,{\bf r}_2) \ . \label{wow}
 \end{equation}
 Remember that
 \begin{eqnarray}
 F_{\alpha\beta}({\bf r}_1,{\bf r}_2)&=&S_{\alpha\beta}
 ({\bf r}_1) + S_{\alpha\beta}({\bf r}_2)-S_{\alpha\beta}
 ({\bf r}_1-{\bf r}_2) \ ,\label{FS}\\
 S_{\alpha\beta}({\bf R})&=& \left<\delta u_{\alpha}({\bf r}+{\bf R},{\bf
 r}) \delta u_{\beta}({\bf r}+{\bf R},{\bf r})\right> \ , \label{Sdef}
 \end{eqnarray}
 where $\delta \bf u$ was defined in (\ref{a1}). Consequently the derivative
 implied in Eq.(\ref{defdel}) picks up only the last contribution in
(\ref{Sdef})
 with the final result which is Eq.(\ref{delta_c}).

 \section{Summary and discussion}
 \label{19-sect7}
 In the paper I in this series we presented a proof of locality in the
 perturbative theory of the velocity structure functions and the velocity
 Green's functions.  This proof excluded the possibility of a perturbative
 mechanism for anomalous scaling behavior of these quantities. In this
 paper we began to explore the nonperturbative  origins of anomalous
 scaling in turbulence. By examining the nonlinear Green's function
 (\ref{c1}) we showed that its diagrammatic series exhibits ladder diagrams
 that produce logarithmic terms that resum to an anomalous power of the
 dimensionless ratio of two separation distances.The results of Section 5
 can summarized as follows:
 \begin{equation}
 \nabla_r^2 G_2(r,r,R,R) \propto r^{-\Delta} \ .
 \label{result}
 \end{equation}
 The analysis of the resummed perturbation series of the nonlinear Green's
 function in section 5 indicated that $\Delta$ has a critical value which
 is $2-\zeta_2$ for which nonperturbative effects may become very important
 and may lead to a renormalization of all the scaling exponents in the
 theory. In Section 6 we demonstrated that $\Delta$ attains exactly this
 critical value.  This result opens up the critical scenario for the
 renormalization of the scaling exponents $\zeta_n$, which will be explored
 in detail in paper III of this series. It will be shown there how the
 critical scenario may result in multiscaling with the outer scale of
 turbulence as the renormalization length.  The deep implications of this
 type of criticality are explained in \cite{95FGLP} and in paper III.

 \acknowledgments
 We are grateful to Daniel Segel and Adrienne Fairhall for
 their critical reading of the manuscript. This work has been
 supported in part by the Naftali Bronicky Fund, the German Israeli
 Foundation and the Basic Research Fund of the Israeli Academy of Sciences.

 \section{Appendix A: Reduction of the Integral Equation}
  Consider the homogeneous part of  Eq.(\ref{e1}), integrate it over
 $t { _{3}}$
 and $t { _{4}}$, and consider it for $t { _{1}}=t { _{2}}=0$, $r { _{1}}=r
 { _{2}}=r$ and $r { _{3}}=r { _{4}}=R$. Dividing the result by $[\tau(r,R)
 G { ^{0}}( {\bf r} { _{0}}|r,R,0 { ^{+}})] { ^{2 }}$ we find the following
 equation:
 \begin{eqnarray}
 \tilde g _{2}(r,R) &=& {1\over [\tau(r,R) G { ^{0}}(
 {\bf r} { _{0}}|r,R,0 { ^{+}})] { ^{2}}}
 \int dx { _{a}}dx { _{b}}dx { _{c}}dx { _{d }}
  \nonumber \\
 &\times& G( {\bf r} { _{0}}|r,r { _{a}},t { _{a}})
  G( {\bf r} { _{0}}|r,r { _{b}},t { _{b}})\Sigma { _{2}}( {\bf r} {
 _{0}}|x { _{a}},x { _{b}},x { _{c}},x { _{d}})
 \nonumber \\
 &\times&\int dt { _{3}}dt { _{4}} G { _{2}}( {\bf r} { _{0}}|x { _{c}},x
 { _{d}},R,t { _{3}},R,t { _{4}})\ .
 \label{e5}
 \end{eqnarray}
 The property of rigidity of $\Sigma { _{2}}$ implies that the main
 contribution to the integral comes from the region $r { _{a}}\sim r {
 _{b}}\sim r { _{c}}\sim r { _{d}}$. Because of the logarithmic situation
 the largest contribution will come from the region of  $r_a$ integration
 in which  $r\ll  r { _{a}} \ll R$. This offers an evaluation of the
 integrations over $r { _{b}}, r { _{c}}$ and $r { _{d}}$ as $r { _{a}} {
 ^{9}}$. We rewrite
 \begin{eqnarray}
 && \!\!\!\!\!\!
 \tilde g _{2}(r,R) = {1\over [\tau(r,R) G { ^{0}}(
 {\bf r} { _{0}}|r,R,0 { ^{+}})] { ^{2}}}
 \int dr { _{a}} r { _{a}} { ^{11 }}
 \nonumber\\
 &\times& \int dt { _{a}}dt { _{b}}dt { _{c}}dt { _{d }}
 G( {\bf r} { _{0}}|r,r { _{a}},t { _{a}})
 G( {\bf r} { _{0}}|r,r { _{a}},t { _{b}})
 \label{e6} \\
 &\times&\Sigma { _{2}}( {\bf r} { _{0}}|r { _{a}},t
 { _{a}}r { _{a}},t { _{b}},r {
 _{a}},t { _{c}},r { _{a}},t { _{d}})
 \nonumber\\
 &\times&
 \int dt { _{3}}dt { _{4}} G { _{2}}( {\bf r} { _{0}}|r { _{a}},t {
 _{c}},r { _{a}},t { _{d}},R,t { _{3}},R,t { _{4}})\ .
 \nonumber
 \end{eqnarray}
 Note now that the typical time scale of $G( {\bf r} { _{0}}|r,r { _{a}},t
 { _{a}})$  is smaller than $\tau(r)$, whereas the typical time scale of
 $\Sigma { _{2}}$ is $\tau(r { _{a}})$ which is much larger than $\tau(r)$.
 Accordingly we can take  $t { _{a}}=t { _{b}}=0$ in the expression for
 ${\Sigma}{ _{2}}$, and perform the $t { _{a}}$ and $t { _{b }}$ integrals
 with the help of the relation (\ref{b12}). The result is
 \begin{eqnarray}
 &&   \!\!\!\!\!\!
 \tilde g _{2}(r,R) = {1\over [\tau(r,R) G { ^{0}}(
 {\bf r} { _{0}}|r,R,0 { ^{+}})] { ^{2}}}
 \int dr { _{a}} r { _{a}} { ^{11 }}
 \nonumber \\
 &\times&\int dt { _{c}}dt { _{d }}[\tau(r,r {
 _{a}}) G( {\bf r} { _{0}}|r,r { _{a}},0 { ^{+}})] { ^{2}}
  \label{e7}\\
 &\times&
  \Sigma { _{2}}( {\bf r} { _{0}}|r { _{a}},t { _{a}}=0,r { _{a}},t {
 _{b}}=0,r { _{a}},t { _{c}},r { _{a}},t { _{d}})
 \nonumber\\
 &\times& \int  dt { _{3}}dt { _{4}} G
 { _{2}}( {\bf r} { _{0}}|r { _{a}},t { _{c}},r { _{a}},t { _{d}},R,t {
 _{3}},R,t { _{4}}) \ .
 \nonumber
 \end{eqnarray}
 To proceed we note that  now $\Sigma { _{2}}$ is determined by one time
 scale, $\tau(r { _{a}})$. On the other hand $G { _{2}}$ depends on three
 time differences, say $t { _{3}}-t { _{c}}$, $t { _{4}}-t { _{c}}$ and $t
 { _{d}}-t { _{c}}$. The time arguments $t { _{c }}$ and  $t { _{d}}$
 cannot exceed $\tau(r { _{a}})$ because of the decay of ${\Sigma}{ _{2}}$.
 For $R \gg r { _{a}}$ this means that  $t { _{3}}-t { _{c }}\approx t {
 _{3}}$, $t { _{4}}-t { _{c }}\approx t { _{4}}$ and $t { _{d}}-t { _{c
 }}\sim \tau(r { _{a}})$. In the evaluation of $G { _{2}}$ we can take $t {
 _{d}}-t { _{c}}$ to be zero.  Using the definition (\ref{e4}) and
 evaluating the integral over $t { _{c}}$ and $t { _{d}}$ as $\tau(r {
 _{a}})$ we have
 \begin{eqnarray}
  &&
 \tilde g_{2}(r,R) = {1\over
 [\tau(r,R) G { ^{0}}( {\bf r} { _{0}}|r,R,0 { ^{+}})] { ^{2}}}
 \int dr { _{a}} r { _{a}} { ^{11 }}
 \nonumber \\
 &\times&
 [\tau(r { _{a}}) \tau(r,r {
 _{a}}) G( {\bf r} { _{0}}|r,r { _{a}},0 { ^{+}})] { ^{2}}
  \label{e8}  \\
 &\times&
 \Sigma { _{2}}( {\bf r} { _{0}}|r { _{a}},t { _{a}}=0,r { _{a}},t {
 _{b}}=0,r { _{a}},t { _{c}}=0,r { _{a}},t { _{d}}=0)
 \nonumber \\
 &\times&
 [\tau(r { _{a}},R) G {
 ^{0}}( {\bf r} { _{0}}|r { _{a}},R,0 { ^{+}})] { ^{2}}\tilde g
 _{2}(r { _{a}},R)\ .
 \nonumber
 \end{eqnarray}
 Using Eqs.(\ref{b13}-\ref{b16}) we find that
 $$\tau(r { _{a}}) \tau(r,r { _{a}})
 \tau(r { _{a}},R)/\tau(r,R) = \big[\tau(r_a)\big]^2 \ .
 $$
 We need now to evaluate $\Sigma { _{2}}$.In the previous
 section we showed that all the diagrams in this series are rigid.
 Consequently the major contribution to the equation (\ref{e1}) comes from
 the regime in which the four coordinates in ${\Sigma} { _{2}}$ are of the
 same order.  Because of this property of rigidity and the scaling
 relations each diagram in the series has the same order of magnitude and
 represents a homogeneous function of its arguments with the same scaling
 index that we denote as ${\sigma}{ _{2}}$. We can evaluate ${\Sigma} {
 _{2}}$ from any one of the diagrams in its series.  For example the first
 diagram with two A vertices gives us $1/r { ^{2}}$ from these vertices,
 $1/(r { ^{6}}( { \tau}(r) { ^{2}})$ from two delta functions $\delta (x {
 _{i}}-x { _{j}})$, and $S { _{2}}(r)$ from the correlator. In total
 \begin{equation}
 \Sigma { _{2}}(r) \sim {S { _{2}}(r)\over r ^8[\tau (r)]^2 } =
 {  C\over r { ^{6}}[\tau (r)]^4 }
 \label{e2}
 \end{equation}
 where we made use of the scaling relation (\ref{b15}), and C is a
 dimensionless coefficient. The last form will be of use in our
 calculations below.  The scaling exponent of $\Sigma { _{2}}(r)$ can be
 read from (\ref{e2}):
 \begin{equation}
 \sigma { _{2}} = 2\zeta { _{2}} - 10 = -6 -4z \ .
 \label{e3}
 \end{equation}
 One can check that this scaling relation, together with the fact that the
 scaling exponent of $G( {\bf r} { _{0}}|x { _{a}},x { _{c}})$ is -3
 guarantees that the second term in the RHS of Eq.(\ref{e1}) has the same
 scaling index as the first term, which is -6.
 Using then the evaluation (\ref{e2}) we find
 \begin{equation}
 \tilde g { _{2}}(r,R) \sim A \int {dr { _{a}} \over
 r { _{a}} } \tilde K (r,r { _{a}},R) \tilde g{_{2}}(r { _{a}},R)
 \label{e9}
 \end{equation}
 where $\tilde K (r,r { _{a}},R)$ is the dimensionless function
 \begin{equation}
 \tilde K (r,r _{a},R) =
 \left[{ r_{a}^{3} G( {\bf r}_{0}|r,r _{a},0^+) G^0
 ( {\bf r}_0|r_a,R,0^+)
 \over
  G^0 ( {\bf r}_0|r,R,0^+)} \right]^{2}\ .
 \label{e10}
 \end{equation}
 When all the scales are in the inertial regime $\tilde K (r,r { _{a}},R)$
 is a function of two ratios only,
 \begin{equation}
 \tilde K (r,r { _{a}},R) = K(y,y { _{a}})\,,\ \  y=r/R\,,
 \ \  y { _{a}} = r { _{a}}/R\  .
 \label{e11}
 \end{equation}
 In this regime we can write the final equations (\ref{e12})-(\ref{e14}).

 \vspace{-.3cm}



 \end{document}